\documentclass[sigconf]{acmart}

\copyrightyear{2023} 
\acmYear{2023} 
\setcopyright{acmlicensed}\acmConference[CCS '23]{Proceedings of the 2023 ACM SIGSAC Conference on Computer and Communications Security}{November 26--30, 2023}{Copenhagen, Denmark}
\acmBooktitle{Proceedings of the 2023 ACM SIGSAC Conference on Computer and Communications Security (CCS '23), November 26--30, 2023, Copenhagen, Denmark}
\acmPrice{15.00}
\acmDOI{10.1145/3576915.3623095}
\acmISBN{979-8-4007-0050-7/23/11}

\usepackage[english]{babel}

\usepackage{amsmath,amstext} 
\usepackage{algorithm}
\usepackage[noend]{algpseudocode}
\usepackage{algpseudocode}
\usepackage[lambda ,advantage, operators, sets , adversary, landau, probability, notions, logic, ff , mm, primitives, events, complexity, asymptotics, keys]{cryptocode}
\usepackage{caption}
\usepackage{booktabs}
\usepackage{tikz}
\usepackage{stmaryrd}
\usepackage{multirow}
\usepackage{xcolor}
\usepackage{xurl}
\usepackage{float}
\usepackage{subfig}
\usepackage{cleveref}

\newcommand{\encrypt}[1]{#1}
\newcommand{\node}[0]{\texttt{node}}

\newcommand{\twoline}[2]{\begin{tabular}{c}#1\\#2\end{tabular}}
\newcommand{\tree}[0]{\mathcal{T}}
\newcommand{\leaves}[0]{\mathcal{L}}
\newcommand{\internal}[0]{\mathcal{D}}
\newcommand{\BB}[0]{\mathbb{B}}

\definecolor{bostonred}{rgb}{0.8, 0.0, 0.0}
\newcommand{\redsymbol}[1]{\textcolor{bostonred}{#1}}

\definecolor{redbg}{RGB}{254,241,240}
\definecolor{redoutline}{RGB}{252,163,152}
\definecolor{redtext}{RGB}{207,24,34}
\definecolor{beigebg}{RGB}{245, 245, 220} %

\DeclareRobustCommand*{\xxcmppdte}[0]{%
  \tikz[baseline=(char.base)]\node[anchor=south west, draw, rectangle, thick, rounded corners=1mm, inner sep=2pt, fill=redbg, draw=redoutline, text=black](char){XXCMP-PDTE};
}

\DeclareRobustCommand*{\xxcmppdtenospace}[0]{%
  \tikz[baseline=(char.base)]\node[anchor=south west, draw, rectangle, thick, rounded corners=1mm, inner sep=2pt, fill=redbg, draw=redoutline, text=black](char){XXCMP-PDTE};%
}

\DeclareRobustCommand*{\rccpdte}[0]{%
  \tikz[baseline=(char.base)]\node[anchor=south west, draw, rectangle, thick, rounded corners=1mm, inner sep=2pt, fill=beigebg, draw=redoutline, text=black](char){RCC-PDTE};
}

\DeclareRobustCommand*{\rccpdtenospace}[0]{%
  \tikz[baseline=(char.base)]\node[anchor=south west, draw, rectangle, thick, rounded corners=1mm, inner sep=2pt, fill=beigebg, draw=redoutline, text=black](char){RCC-PDTE};%
}

\begin{document}

\title{Level Up: Private Non-Interactive Decision Tree Evaluation using Levelled Homomorphic Encryption}


\author{Rasoul Akhavan Mahdavi}
\email{rasoul.akhavan.mahdavi@uwaterloo.ca}
\affiliation{%
  \institution{University of Waterloo}
  \city{Waterloo}
  \country{Canada}
}

\author{Haoyan Ni}
\email{h27ni@uwaterloo.ca}
\affiliation{%
  \institution{University of Waterloo}
  \city{Waterloo}
  \country{Canada}
}

\author{Dimitry Linkov}
\email{dimitry.linkov@uwaterloo.ca}
\affiliation{%
  \institution{University of Waterloo}
  \city{Waterloo}
  \country{Canada}
}

\author{Florian Kerschbaum}
\email{florian.kerschbaum@uwaterloo.ca}
\affiliation{%
  \institution{University of Waterloo}
  \city{Waterloo}
  \country{Canada}
}

\renewcommand{\shortauthors}{Mahdavi et al.}

\begin{abstract}

As machine learning as a service continues gaining popularity, concerns about privacy and intellectual property arise. Users often hesitate to disclose their private information to obtain a service, while service providers aim to protect their proprietary models. Decision trees, a widely used machine learning model, are favoured for their simplicity, interpretability, and ease of training. In this context, Private Decision Tree Evaluation (PDTE) enables a server holding a private decision tree to provide predictions based on a client's private attributes.
The protocol is such that the server learns nothing about the client's private attributes. Similarly, the client learns nothing about the server's model besides the prediction and some hyperparameters.

In this paper, we propose two novel non-interactive PDTE protocols, \xxcmppdte and \rccpdtenospace, based on two new non-interactive comparison protocols, XXCMP and RCC. Our evaluation of these comparison operators demonstrates that our proposed constructions can efficiently evaluate high-precision numbers. Specifically, RCC can compare 32-bit numbers in under 10 milliseconds.

We assess our proposed PDTE protocols on decision trees trained over UCI datasets and compare our results with existing work in the field. Moreover, we evaluate synthetic decision trees to showcase scalability, revealing that \rccpdte can evaluate a decision tree with over 1000 nodes and 16 bits of precision in under 2 seconds. In contrast, the current state-of-the-art requires over 10 seconds to evaluate such a tree with only 11 bits of precision.

\end{abstract}



\maketitle

\section{Introduction}
With the widespread adoption of machine learning (ML) in many industries, there is a growing interest to offer cloud-based machine learning services~\cite{aws, azure, bigml, google}.
However, using cloud-based ML services necessitates clients to share their confidential data with providers to benefit from these services.
For many users, this prerequisite raises serious concerns about the potential loss of privacy.
Additionally, companies which wish to collaborate and use each other's services cannot risk exposing their customers' and employees' data. This creates a barrier to potential business collaborations.
Moreover, service providers are unwilling to relinquish classification models to users, which could eliminate their competitive advantage and put the users in the training data at risk.

Decision trees are a well-known ML algorithm which are still used widely in many tasks due to their simplicity, interpretability, and ease of training~\cite{quinlan1986induction, hastie2009elements}.
Private Decision Tree Evaluation (PDTE) is a protocol for providing a prediction using a private decision tree hosted by a server on a private input provided by a client.
At the end of the protocol, the server learns nothing about the client's input (input privacy), and the client learns nothing about the server's decision tree (model privacy) other than the result of the inference and some hyperparameters.

One set of solutions is interactive, where the client and server exchange messages~\cite{tueno, Tai2017PrivacyPreservingDT, bost2014machine, bai2022} in multiple rounds. These solutions are based on tools such as multi-party computation, secret-sharing and garbled circuits~\cite{kiss2019}.
Another category of solutions is non-interactive approaches, where the client can submit the query and go offline until the response is ready. This is great for the setting where the client lacks computational power or suffers from limited bandwidth. All existing solutions to non-interactive PDTE use levelled or fully homomorphic encryption~\cite{xcmp, sortinghats, Tueno2019NonInteractivePD, akavia2022}.
Solutions requiring levelled homomorphic encryption select parameters that depend on the comparison's precision. This limits the scalability of the solution significantly~\cite{tueno, xcmp}. Solutions using fully homomorphic encryption do not suffer the same issue, but individual operations are inefficient due to the expensive bootstrapping procedure. The efficiency problem is exacerbated because some fully homomorphic schemes do not support SIMD operations, so inferring multiple samples in parallel is not possible.

This work proposes two non-interactive PDTE protocols using levelled homomorphic encryption schemes.
We propose a protocol where the multiplicative depth of the entire PDTE protocol does not depend on the tree's structure or the number of client attributes, making it more scalable and practical. Our solution can scale to trees with over 1000 decision nodes and 100 client attributes.
At the heart of our proposed protocols are efficient non-interactive comparison operators. First, an extension of XCMP by Lu et al.~\cite{xcmp}, which we denote as Extended XCMP (XXCMP). This extension supports arbitrary precision and is implemented using automorphisms in the SEAL library.
Second, is a comparison operator based on the concept of range covers~\cite{range-cover, shirangequery}, combined with constant-weight equality operators proposed by Mahdavi and Kerschbaum~\cite{cpir}. We denote this comparison operator as Range Cover Comparison (RCC).
Both operators are implemented with FV, an RLWE-based cryptosystem, but in different modes of operation~\cite{fan2012somewhat}.
Comparison operators are the core building block in all non-interactive PDTE protocols~\cite{xcmp, akavia2022, probonite, tueno, sortinghats}.
Our proposed comparison operators can efficiently compare numbers with arbitrary precision. In contrast, previous work is limited in precision due to efficiency.
XCMP is limited to 13 bits, while the operator used by Cong et al.~\cite{sortinghats} can only compare 11-bit numbers.
The comparison operator of Tueno et al.~\cite{Tueno2019NonInteractivePD} can compare numbers with arbitrary precision. However, the parameters of the levelled FHE scheme grow with bit precision, limiting the solution's efficiency for high-precision inputs.
Our evaluation shows that XXCMP and RCC are up to 100 times faster than the operators proposed by Tueno et al.
Since our comparison operators perform comparisons with arbitrary precision, models such as decision trees need not be retrained with low precision to enable private inference.

We use the SumPath algorithm described in ~\Cref{sec:tree-traversal} to evaluate the paths in the decision tree. SumPath requires no multiplications and hence does not increase the multiplicative depth of the circuit.
By combining SumPath with XXCMP and RCC, we propose two new PDTE protocols, \xxcmppdte and \rccpdtenospace, which use RLWE-based cryptosystems in two different modes of operation.
While using RLWE-based cryptosystems with SIMD support has previously been proposed~\cite{Tueno2019NonInteractivePD}, it was only to infer multiple samples in parallel and reduce amortized time. In contrast, our work uses SIMD operations to not only perform multiple inferences but also to speed up even a single inference, which reduces client latency. This approach also allows the client to pack more information into fewer ciphertexts, reducing the overall communication between the client and server when only one inference is performed.

In our evaluation, we train decision trees with different bit precision over commonly used UCI datasets~\cite{Dua:2019}.
Our evaluation shows that \xxcmppdte and \rccpdte are up to 5 times faster than SortingHats~\cite{sortinghats}, which is a state-of-the-art solution for PDTE.
This advantage increases when inferring many samples in parallel.
The performance of PDTE, i.e., the communication and computation cost, depends mainly on three factors: precision, the number of decision nodes, and the number of client attributes. We perform an ablation study over these parameters using synthetic decision trees to demonstrate the dependency between these parameters and the performance.
Our experiments show that the number of client attributes affects communication and the number of decision nodes affects computation, while bit precision influences both metrics.
Moreover, our experiments show that \rccpdte is the most scalable solution. It outperforms all other solutions when the number of decision nodes grows, and the number of client attributes increases. Specifically, it can infer decision trees with over 1000 decision nodes and 16-bit precision in under 2 seconds. SortingHats requires more than 10 seconds to evaluate such a decision tree with only 11 bits of precision.

In summary, our contributions are as follows:
\begin{itemize}\itemsep0mm
    \item Two non-interactive comparison protocols which we denote as XXCMP and RCC, that can compare numbers with arbitrary precision using levelled homomorphic encryption.
    \item Two non-interactive PDTE protocols \xxcmppdte and \rccpdtenospace.
    \item Evaluation of proposed comparison operators with state-of-the-art protocols.
    \item Ablation of PDTE over the number of client attributes and the size of the decision tree, which shows \rccpdte to be the most scalable solution. Our experiments show that \rccpdte can evaluate decision trees with up to 1000 nodes in less than 2 seconds.
\end{itemize}

\begin{table*}
    \centering
    \begin{tabular}{c|c|c|c|c}
    \toprule
        \textbf{Encoding} & \textbf{Plaintext Space} & 
        \multicolumn{3}{c}{\textbf{Operation}}
        \\
    \midrule
        \multirow{4}{*}{Polynomial} & \multirow{4}{*}{$R_p = \frac{\ZZ_p[X]}{X^N+1}$}
              & Polynomial Add & $c_1(x), c_2(x)\in \mathcal{C}$ & $m_1(x)+m_2(x)$ \\
            & & Plain Polynomial Mult. & $c_1(x)\in \mathcal{C}$, $m_2(x)\in R_p ~~~$ & $m_1(x)m_2(x)$ \\
            & & Polynomial Mult. & $c_1(x), c_2(x)\in \mathcal{C}$ & $m_1(x) m_2(x)$ \\
            & & Oblivion Expansion~\cite{sealpir,chen2019onion} & $c_1(x)\in \mathcal{C}, k\in \NN\cup\{0\}$ & Coefficient of $x^k$ in $m_1(x)$\\
        \midrule
        \multirow{4}{*}{Batched} & \multirow{4}{*}{$\ZZ_p^{N/2}$} & 
            SIMD Add & $c_1(x), c_2(x)\in \mathcal{C}$ & $m_1(x)\oplus m_2(x)$ \\
            & & Plain SIMD Mult. & $c_1(x)\in \mathcal{C}$, $m_2(x)\in \ZZ_p^{N/2}$ & $m_1(x)\otimes m_2(x)$ \\
            & & SIMD Mult. & $c_1(x), c_2(x)\in \mathcal{C}$ & $m_1(x) \otimes m_2(x)$ \\
            & & Circular (Right) Rotation & $c_1(x)\in \mathcal{C}, k\in \NN$ & $\text{Rotate}_k(m_1(x))$ \\
    \bottomrule
    \end{tabular}
    \caption{Different encodings for the FV cryptosystem. In all operations, we have $c_1(x), c_2(x)\in \mathcal{C}$ that encrypt $m_1(x), m_2(x)$, respectively.
    Operations over plaintext polynomials happen in $R_p$.
    $\oplus$ and $\otimes$ denote element-wise addition and multiplication modulo $p$ between two vectors. Rotate$_k$ denotes the circular right rotation of a vector by $k$ slots.}
    \label{tab:fv-encoding}
\end{table*}

In \Cref{sec:background}, we review necessary background material such as homomorphic encryption, decision trees and private decision tree evaluation, range covers and tree traversal algorithms.
We also compare the properties of related work on non-interactive comparison and non-interactive PDTE in this section.
We describe our constructions, XXCMP and RCC, two non-interactive private comparison protocols in \Cref{sec:non-interactive-cmp-eval}. In \Cref{sec:levelled-pdte-constructs}, we describe our PDTE protocols, \xxcmppdte and \rccpdtenospace. 
In \Cref{sec:pdte-eval}, we compare XXCMP and RCC with other non-interactive comparison protocols and then compare PDTE protocols with our constructions.
We conclude in \Cref{sec:discussion} with a discussion on limitations and future work.

\begin{table}[b]
    \centering
    \begin{tabular}{cl}
    \toprule
        Symbol & Description \\
    \midrule
        $\lambda$ & Security parameter \\
        $N$ & Polynomial modulus degree\\
        $p$ & Plaintext modulus \\
        $\redsymbol{x}$ & Encryption of $x$ \\
\midrule
        $\mathcal{M}$ & $(\tree, \textbf{a}, \textbf{t}, \textbf{v})$ \\
         $\tree$ & Decision tree nodes \\
         $\internal$ & Set of internal decision nodes in $\tree$ \\
         $\leaves$ & Set of leaf nodes in $\tree$ \\
         $m$ & Number of decision nodes ($|\internal|$) \\
         $\textbf{x}$ & Client attribute vector \\
         $\textbf{a}$ & Node to attribute mapping \\
         $\textbf{t}$ & Threshold value function \\
         $\textbf{v}$ & Leaf value function \\
         $d$ & Depth of the decision tree \\
         $k$ & Number of classification labels \\
\midrule
        $h$ & Hamming weight \\
        $\ell$ & Constant-weight code length \\
        $n$ & Bit Precision \\
        $[n]$ & $\{0,1,\cdots,n-1\}$ \\
        $\BB$ & $\{0,1\}$ \\
    \bottomrule
    \end{tabular}
    \caption{Summary of notation}
    \label{tab:notation}
\end{table}

\section{Background \& Related Work}
\label{sec:background}

\Cref{tab:notation} summarizes the notation used in the background section and throughout the paper. Throughout the paper, we use $[n]$ to refer to the set $\{0,1,\cdots,n-1\}$, for $n\in\NN$.

\subsection{Homomorphic Encryption}
Homomorphic Encryption (HE) is a form of encryption that permits computation on the data while in encrypted form.
Levelled homomorphic encryption (LHE) schemes permit computation of circuits with a limited multiplicative depth~\cite{fan2012somewhat, bgv}. The parameters of the cryptosystem are chosen based on the multiplicative depth.
Hence, we try to design algorithms with a lower multiplicative depth to enhance performance. 
A fully homomorphic encryption (FHE) scheme permits an unlimited amount of operations with the help of bootstrapping~\cite{ducas2015fhew, chillotti2020tfhe}. However, FHE is more computationally expensive and requires large cryptographic keys for setup.

\paragraph{Fan–Vercauteren (FV) Cryptosystem.}
The Fan–Vercauteren (FV) cryptosystem~\cite{fan2012somewhat} is a lattice-based homomorphic cryptosystem.
An FV ciphertext is an array of polynomials, each from $R_q = \ZZ_q[X]/(X^N+1)$, where $q$ is called the \emph{coefficient modulus}. In the simplest case, the ciphertext is only two polynomials. Let $\mathcal{C}$ denote the ciphertext space. $N$ and $q$ determine both the security parameter and how many homomorphic operations can be performed on ciphertexts before decryption is necessary.
Inputs in this cryptosystem can be encoded in two formats. \Cref{tab:fv-encoding} shows the two encoding types and the corresponding operations that can be performed.

Microsoft SEAL~\cite{sealcrypto} implements the FV cryptosystem and supports all the operations mentioned in \Cref{tab:fv-encoding}. We use the SEAL library for our implementations in this work.

\subsection{Non-interactive Private Comparison}
\label{sec:background-non-interactive-cmp}

A comparison operator is a function $f:D\mapsto\{0,1\}$ such that for $x,y\in D$,
\begin{align}
    f(x,y) = \mathbb{I}[x \leq y]
\end{align}

A private comparison is a protocol where two inputs, $x$ and $y$, are provided, such that one or both of them are encrypted. The output of the protocol is $f(x,y)$ in encrypted form.

In this work, we are particularly interested in non-interactive solutions to this problem. This is useful in protocols where the encrypted input is provided by a lightweight client which may go offline after providing the input. Below we describe three non-interactive private comparison protocols from the literature.

\paragraph{Folklore Private Comparison}

The folklore comparison algorithm compares two $n$-bit numbers in binary format. This is identical to how binary numbers are compared in the clear, with some adaptations to make it easier to compute using HE.
The multiplicative depth of the algorithm is $1 + \log_2 (n+1)$, which poses a burden to compute efficiently using HE. \Cref{alg:folklore-cmp} shows this algorithm. The inputs are binary vectors and all operations are element-wise. 
We also use the $\textsc{RightShift}_k$ function, which logically shifts the contents of any vector or bitstring by $k$ positions.
Additions and multiplications can also be replaced with XOR and AND operations.

\begin{algorithm}[ht]
	 \caption{\textsc{Folklore Comparison}($\mathbb{I}[a \leq b]$)}
	 \label{alg:folklore-cmp}
	 \begin{flushleft}
		 \textbf{Input:} $a,b\in\{0,1\}^n$
	 \end{flushleft}
	 \begin{algorithmic}[1]
        \State $\theta_{eq} \leftarrow 1 - (a-b)^2$
        \State $\theta_{gt} \leftarrow (1-a) \cdot b$
        \State $ \theta_{\text{PrefixEq}} = \theta_{gt} \cdot \prod_{k=0}^{n-1} \texttt{RightShift}_k(\theta_{eq}) $
        \vspace{1mm}
        \State $\theta \leftarrow \sum_{i} \theta_{\text{PrefixEq}}[i]$
	 \end{algorithmic}
	 \begin{flushleft}
    	 \textbf{Output:} $\theta \in \{0,1\}$ \\
	 \end{flushleft}
\end{algorithm}

Variations of folklore comparison have been implemented in many works using levelled homomorphic encryption~\cite{Tueno2019NonInteractivePD, hao2023privacy}.

\paragraph{XCMP}
Lu et al.~\cite{xcmp} proposed a comparison operator called XCMP which compares two encrypted numbers using levelled HE.
Cong et al. introduced a variation using TFHE where one operand is unencrypted~\cite{sortinghats}.
This relaxation reduces the runtime of the comparison.
\Cref{alg:xcmp} shows this variant of XCMP which compares an encrypted input, $a$, provided by the client with an unencrypted $b$ provided by the server. The output is $\mathbb{I}[a\leq b]$.
Inputs are encoded as RLWE ciphertexts, i.e., Polynomial encoding from \Cref{tab:fv-encoding}. 

\begin{algorithm}[ht]
	 \caption{\textsc{XCMP}($\mathbb{I}[a\leq b]$) \cite{xcmp}}
	 \label{alg:xcmp}
    \begin{algorithmic}[1]
        \Procedure{XCMP}{$A,b$}
        \Comment{$A=X^a, a,b\in [N]$}
            \State $T \leftarrow \frac{1}{2}X^{-b}\cdot (1 + X + \cdots + X^{N-1})$
            \State $R \xleftarrow{\$} R_p$ and $R[0] = 1/2 \mod p$
            \vspace{1mm}
            \State $\encrypt{C} = A \cdot T + R$
            
            \Return $C$
        \EndProcedure  
    \end{algorithmic}
\end{algorithm}

The result of the comparison is in the constant term of $C$.

\paragraph{Iliashenko and Zucca~\cite{iliashenko2021faster}}
The comparison function of $x,y\in \ZZ_p$ can be represented as either a bivariate polynomial of the two inputs or a univariate function of the difference.
Iliashenko and Zucca showed how to exploit the structure of these polynomials to efficiently evaluate the comparison function.
Their main observation was that these polynomials have many zero coefficients which can be ignored.

Based on this observation, they showed that comparison using FHE schemes that operate over arithmetic circuits can be efficient. 

\subsection{Decision Trees and PDTE}

\paragraph{Decision Trees.}
A decision tree is a classification algorithm which classifies input data by sequentially checking a series of criteria. The simplest form of a decision tree is represented by a binary tree where each internal node compares an attribute with a threshold.
Each leaf is assigned a classification value (or simply a class).
To classify a data point, we start at the root of the tree, perform a comparison and move to the right or left child, depending on the result of the comparison.
We continue until we reach a leaf and output the class of that leaf.

More formally, a decision tree consists of a set of nodes $\tree = \mathcal{D} \cup \mathcal{L}$, where $\mathcal{D}$ and $\mathcal{L}$ are the set of decision nodes and leaf nodes, respectively.
There also exists an attribute vector, $\textbf{x}$, and three functions:
\begin{itemize}\itemsep0mm
    \item $\textbf{a}:\mathcal{D} \mapsto [|\textbf{x}|]$ maps decision nodes to attribute indices.
    \item $\textbf{t}:\mathcal{D} \mapsto \ZZ$ maps internal nodes to threshold values.
    \item $\textbf{v}:\mathcal{L} \mapsto \ZZ$ maps leaf nodes to classification values.
\end{itemize}

We denote the tuple $\mathcal{M}=(\tree, \textbf{a}, \textbf{t}, \textbf{v})$ as the decision tree model.

\paragraph{Private Decision Tree Evaluation (PDTE)}
Private Decision Tree Evaluation (PDTE) is a protocol between a server and a client where the server holds the model, $\mathcal{M}$, and the client holds the attribute vector, $\textbf{x}$.
The goal is to infer the tree on the client's attribute vector such that the server does not learn anything about the client's input.
Moreover, the client should not learn anything about the server's private decision tree other than the classification result and some hyperparameters.

The client may try to steal the model with black-box access to an API through carefully crafted queries to the server~\cite{tramer2016stealing, papernot2017practical, shokri2017membership, juuti2019prada,10.1145/3274694.3274740}.
Such an attack is outside the scope of this work, but defences against such attacks is an active area of research~\cite{tramer2016stealing, juuti2019prada}.

\subsection{Tree Traversal}
\label{sec:tree-traversal}

\begin{figure}[t]
    \centering
    \includegraphics[width=\columnwidth]{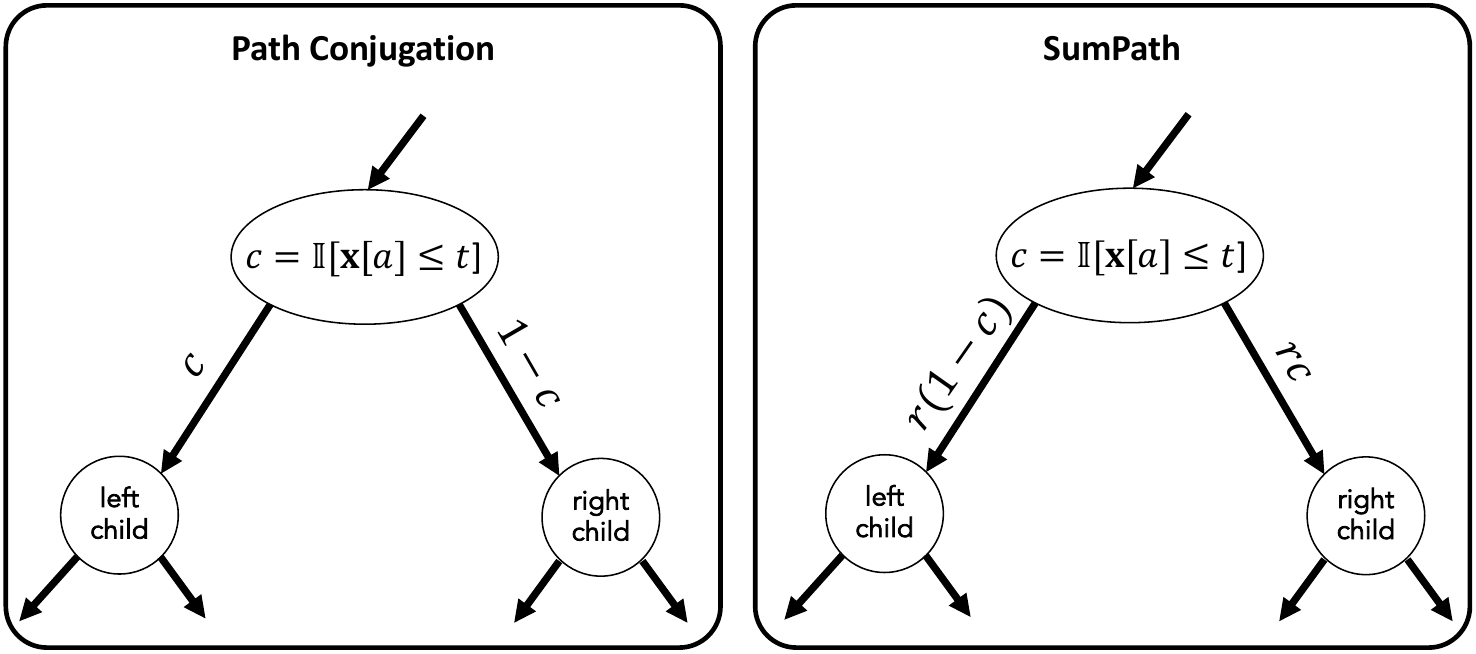}
    \caption{
        Labelling of edges in two tree traversal methods, Path Conjugation and SumPath.
        $\textbf{x}$ is the client attribute vector.
        $a$, and $t$ are the attribute index and threshold for the parent node, respectively.
        In SumPath, $r$ is either a random number or 1.
    }
    \label{fig:node-labels}
\end{figure}

One of the main steps in all private decision tree evaluation protocols is traversing the tree from the root to the leaves, as discussed by Kiss et al.~\cite{kiss2019}.
We denote the leaf holding the result of the prediction as the \emph{result leaf} and the value of that leaf as the \emph{result value}.
If all comparisons in the decision nodes are computed as a binary output, there are two methods to aggregate the results and compute the result leaf and value.
We denote these two methods as \emph{Path Conjugation} and \emph{SumPath}.
For each method, the edges in a tree are annotated as shown in \Cref{fig:node-labels}.

In Path Conjugation, for each leaf in the tree, all values on the edges between the root and that leaf are conjugated. Consequently, the result leaf will have a value of one, and all other leaves will have a value of zero. This process can be computationally optimized by reusing computations in inner nodes~\cite{Tueno2019NonInteractivePD}. Some works, such as those by Sortinghats and Tueno, utilize this approach with fully homomorphic encryption and the CMux gate in TFHE~\cite{sortinghats, tueno}. However, this approach is not practical when using levelled homomorphic encryption, given that the multiplicative depth of the circuit will depend on the depth of the tree.

On the other hand, the SumPath method has been proposed previously for use with additive encryption~\cite{Tai2017PrivacyPreservingDT, kiss2019} or levelled homomorphic encryption in an arithmetic field~\cite{tueno}.
Edges are annotated as indicated in \Cref{fig:node-labels}. For each leaf in the tree, the sum of all the edges in the tree from the root to that leaf is assigned to that leaf. Only the result leaf will have a sum of zero, and all others will be non-zero.
By returning the sum on each leaf, plus the masked value on each leaf, the client can infer the correct result value.
The advantage of this approach is that no multiplications or conjugations are required, which makes it computationally inexpensive and does not increase the multiplicative depth.

\subsection{Range Covers \& Point Encoding}
Range covers are a method to represent intervals in a manner that is concise and easy to use in secure computation~\cite{range-cover, shirangequery}.
Let $T$ be a binary tree of internal nodes of prefixes numbers in $[2^n]$ and leaf nodes of the elements in $[ 2^n]$.
The children of a prefix $p$ are $p|0$ and $p|1$, and the root is the empty prefix.
A \emph{range cover} $RC_n(a,b)$ is a set of nodes in $T$, such that its set of children at the leaf level are all elements in the range $[a,b]$, where $a,b\in [2^n]$.
The \emph{best range cover} contains at most $2 \log n$ nodes with at most $2$ nodes at each level of $T$ (excluding the root level). 
A uniform range cover is a modification of the best range cover such that each level of the tree contains exactly 2 nodes~\cite{range-cover, shirangequery}.
This is achieved by padding the best range cover with dummy nodes at all levels if fewer than two nodes are chosen.
The advantage of a uniform range cover is that the size is independent of the interval, which is a requirement in private protocols.
For every range $[a,b]$ there exists a best range cover and uniform range cover.

A \emph{point encoding} $PE_n(c)$ of an element $c \in [2^n]$ is the set of nodes from the leaf $c$ to the root (except the root itself).
A point encoding consists of $n$ nodes with one node at each level of $T$ (except the root level).
We denote as $RC_n(a,b)[i]$ and $PE_n(c)[i]$ the $i$-th element of a range cover or encoding, respectively.

We can test the relationship $c \in [a,b]$ given $RC_n(a,b)$ and $PE_n(c)$ 
by checking to see if there are any common prefix nodes between the range cover and point encoding. We give the full algorithm on how to do this check in \Cref{alg:rc-compare} in the appendix.

Figure~\ref{fig:rangecover} shows an example: the prefix tree for $[0,2^3-1]$ with the range cover for $[0,4]$, indicated by the shaded boxes, and point encodings for $2$ and $6$.
We can see that $RC_3(0,4)$ and $PE_3(2)$ have the prefix node corresponding to $0$ in common whereas $RC_3(0,4)$ and $PE_3(6)$ have no prefix node in common.

\begin{figure}[htb]
    \centering
    \includegraphics[width=\columnwidth]{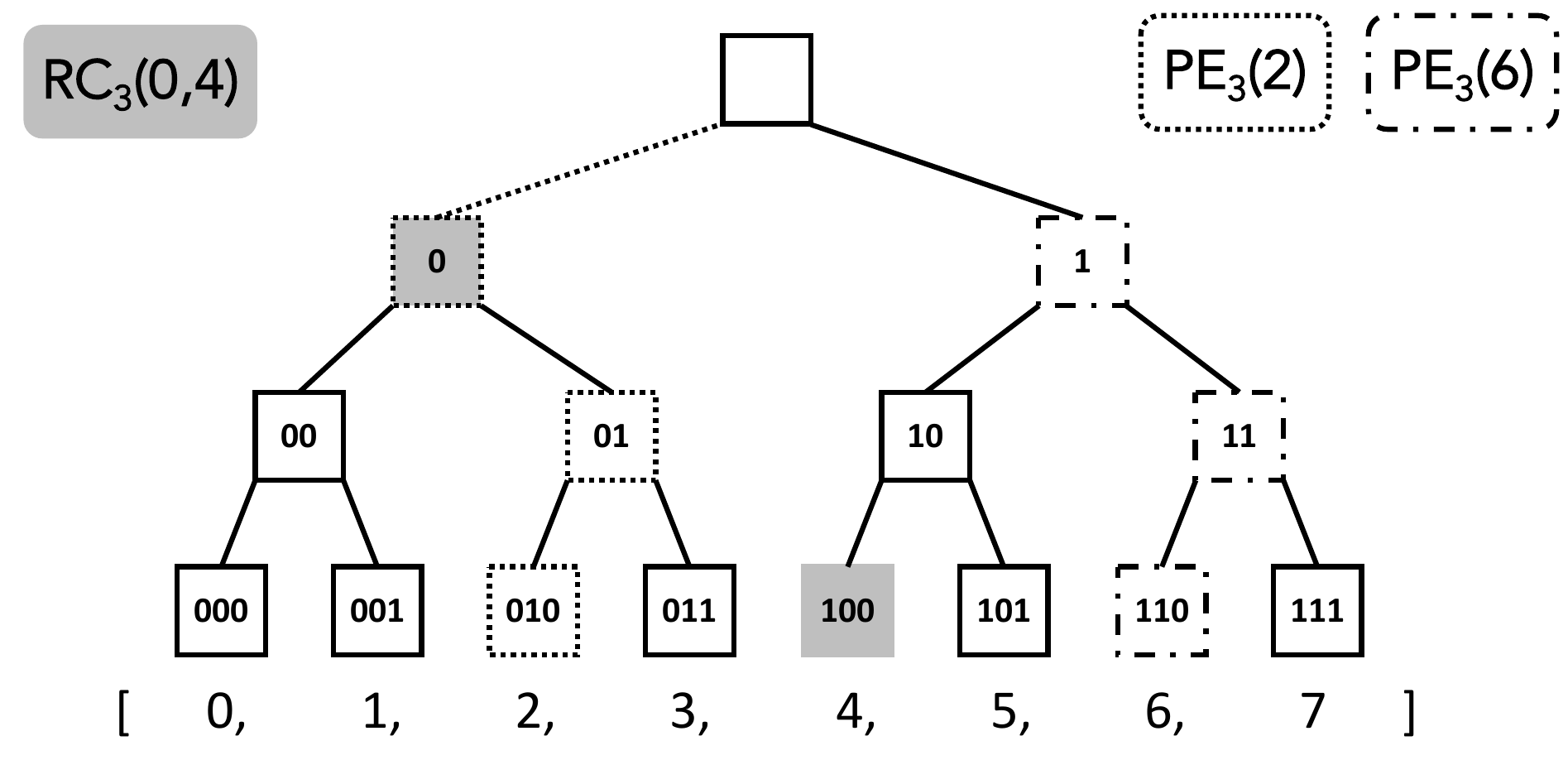}
    \caption{Range cover for $[0,4]$ (shaded boxes) and point encoding of $2$ and $6$ in domain $[0,2^3-1]$. In this example, $RC_3(0,4)$ and $PE_3(2)$ have a prefix node in common, given that $2\in [0,4]$ but $RC_3(0,4)$ and $PE_3(6)$ have no prefix node in common since $6 \notin [0,4]$.}
    \label{fig:rangecover}
\end{figure}

\subsection{Constant-weight Equality Operators}

Constant-weight equality operators, proposed by Mahdavi and Kerschbaum, are equality operators with a constant multiplicative depth, independent of the bitlength of the operands~\cite{cpir}.
To use the constant-weight equality operator, numbers are represented as constant-weight codes with Hamming weight $h$.
We make an adjustment to the encoding algorithm to encode null values as well. Null values are encoded as the all-zero string.
We include the adjusted algorithm from the work of Mahdavi and Kerschbaum in \Cref{sec:cw-stuff}.
To compare constant-weight codewords, we use the arithmetic constant-weight equality operator in this work, which has a multiplicative depth of $1 + \log_2 h$.
We use this operator to be able to compare many pairs of numbers simultaneously, and in a SIMD fashion.
This operator is shown in \Cref{alg:arith-cw-eq}.

\begin{algorithm}[ht]
     \caption[]{Arithmetic Constant-weight Equality Operator~\cite{cpir}}
     \label{alg:arith-cw-eq}
     \begin{algorithmic}[1]
        \Procedure{Arith-CW-Eq-Op}{$x, y$} \Comment{$x, y \in CW(\ell, h)\cup \{0^\ell\}$}
            \vspace{1mm}
            \State $h' = \sum_{i\in[\ell]}x[i] \cdot y[i]$
            \vspace{2mm}
            \State $e = 1/h! \cdot \prod_{i\in[h]} (h' - i)$
            \vspace{1mm}
        \EndProcedure
        \Return $e$ \Comment{$e \in \BB$}
     \end{algorithmic}
\end{algorithm}

\section{Related Work on PDTE}

\subsection{Interactive PDTE}

Decision trees can be privately evaluated using two-party computation (2PC) or using a combination of 2PC and homomorphic encryption. There are also solutions which use tools such as secret sharing, garbled circuits, and oblivious transfer~\cite{brickell2007, lu2023, bai2022, kiss2019}.
Unlike homomorphic encryption, 2PC is usually communication-bound, i.e., reducing the communication complexity or rounds is necessary to increase efficiency.
However, since this work focuses on non-interactive decision tree evaluation, we provide only a few selected examples of the work in private decision trees using 2PC.

2PC, e.g., using Yao's garbled circuit, implements a constant-round protocol.
Brickell et al.~\cite{brickell2007} present an improved constant-round protocol using 2PC and homomorphic encryption.
Kiss et al.~\cite{kiss2019} later surveyed this and a number of other protocols and identified several design options and new combinations of these protocols systematizing the protocols.

However, 2PC usually leads to communication complexity which is exponential in the tree depth since all branches need to be evaluated.
Using different techniques, one can also reduce the communication complexity.
Tueno et al.~\cite{tueno} propose to use oblivious RAM (ORAM).
Bai et al.~\cite{bai2022} propose to use additive homomorphic encryption or pseudo-random functions.

Given enough network capacity, it is possible to train decision tree classifiers using 2PC or in general multi-party computation (MPC).
Lindell and Pinkas \cite{lindell2000} present a theoretic design of a 2PC protocol for computing the logarithm in the ID3 training algorithm.
Since then, many proposals for practical systems have been made, e.g., by Wu et al.~\cite{wu2020}, by Zheng et al.~\cite{zheng2021} and by Lu et al.~\cite{lu2023}.

\begin{table*}[t]
    \centering
    \begin{tabular}{c|c|c|c|c|c|c}\toprule
                            & PROBONITE~\cite{probonite} & PDT-Bin~\cite{Tueno2019NonInteractivePD} & PDT-Int~\cite{Tueno2019NonInteractivePD}  & SortingHats~\cite{sortinghats} & \xxcmppdte & \rccpdte \\ \midrule
        Supports Unbalanced  & $\times$ & \checkmark & \checkmark & \checkmark & \checkmark & \checkmark \\
        Attribute Selection  & PIR & Clear & Clear & Clear & Clear & Clear \\
        Comparison           & PBS (CT-CT
    ) & Folklore & Lin-Tzeng~\cite{lin-tzeng} & XCMP-CT-PT & \textbf{XXCMP} & \textbf{RCC} \\
        Path Evaluation      & CMux & AND & SumPath & CMux & SumPath & SumPath \\
        Batchable            & $\times$ & \checkmark & \checkmark & $\times$ & \checkmark & \checkmark \\ 
        Bit Precision            & $< 8^*$ & $n$ & $n$ & 11 & $n$ & $n$ \\
        \midrule
        Levelled or FHE      & FHE & LHE or FHE & LHE(BGV) & FHE(TFHE) & LHE(FV) & LHE(FV) \\
        \# of Comparisons    & $O(d)$ & $|\internal|$ & $|\internal|$ & $|\internal|$ & $|\internal|$ & $|\internal|$ \\
        Query Complexity     & $O(|\textbf{x}|)$ & $O(n |\textbf{x}|)$ & $O(n |\textbf{x}|)$ & $O(N|\textbf{x}|)$ & $O(N|\textbf{x}|)$ & $O(\ell n |\textbf{x}|)$ \\
        Mult. Depth          & N/A & $\log_2 n$ & $\log_2 n$ & N/A & $\ceil{\log_2(n/\log_2 N)}$ & $1+\log_2 h$ \\
        \bottomrule
    \end{tabular}
    \caption{
    Properties of Non-interactive Private Decision Tree Evaluation Protocols.
    $a$ is the number of client attributes, $d$ is the depth of the tree, $\internal$ is the set of internal decision nodes, and $\textbf{x}$ is the client attribute vector.
    * The precision of PROBONITE depends on the choice of parameters for the LWE scheme but is typically less than 8 bits.}
    \label{tab:compare-pdte}
\end{table*}

\subsection{Non-Interactive PDTE}
Interactive protocols are great for computational efficiency but usually require high network activity.
Moreover, they require several rounds of interaction between the client and server.
This is not a good solution if the client is not strong or has a limited network connection.
The client may wish to go offline while waiting for the result in many use cases.
Non-interactive approaches to private decision tree evaluation primarily use homomorphic encryption.
Some works used levelled homomorphic encryption~\cite{xcmp, tueno, akavia2022} while others require a fully homomorphic scheme~\cite{tueno, sortinghats, probonite}.
A decision tree is comprised of many comparisons between client attributes and thresholds, so at the heart of all protocols is an efficient comparison operation.
The primary approach in many works is to reduce the precision of the comparison to make it efficient~\cite{xcmp, sortinghats}.
For this to be possible, the decision tree evaluation has to be fine-tuned for low precision. Quantization-aware training is one approach to this.
If a model already exists with high precision, it can not be used as it is and has to be retrained.
\Cref{tab:compare-pdte} summarizes the properties of related work.
We also provide a short description of how each protocol works.

\paragraph{XCMP PDTE~\cite{xcmp}}
Lu et al. proposed a non-interactive PDTE protocol based on XCMP, described in \Cref{sec:background-non-interactive-cmp}. 
All comparisons are performed using XCMP, and all paths are evaluated using the same method as the work of Tai et al.~\cite{Tai2017PrivacyPreservingDT}.
This protocol's main limitation is that the precision of the comparison which they use is limited to 13 bits, i.e., $\log_2 N$.
The protocol's design was such that the model holder could also encrypt the thresholds of the model and delegate the evaluation to a third-party server. In this case, the third party would learn the tree's structure but not the thresholds.
    
\paragraph{PDT-Bin \& PDT-Int~\cite{Tueno2019NonInteractivePD}}

Tueno, Boev, and Kerschbaum proposed two PDTE protocols using a binary and arithmetic circuit.
In their first construction, denoted as PDT-Bin,
In this construction, numbers are represented using binary encoding and compared using a Folklore comparison.
Traversal of the decision tree in PDT-Bin is performed using homomorphic AND operations, ultimately yielding a single result that represents the outcome of the classification.

In addition to PDT-Bin, Tueno, Boev, and Kerschbaum introduced another protocol, PDT-Int, which is based on levelled homomorphic encryption, specifically employing the BGV scheme. For the comparison operator in PDT-Int, they adopted a variation of the Lin-Tzeng~\cite{lin-tzeng} protocol, which outputs zero in the case of a match and a random number otherwise. To traverse paths in the decision tree, the SumPath technique is employed, and one value corresponding to each leaf node is returned.

\paragraph{PROBONITE~\cite{probonite}}
Azogagh et al. proposed a PDTE protocol named PROBONITE which evaluates only one path of the tree~\cite{probonite}. 
In contrast to other non-interactive protocols, which perform one comparison for each node in the decision tree and evaluate all paths, PROBONITE only evaluates one path. 
The protocol starts at the root and traverses one path down the tree. 
This is done by using two subprocedures: 1) \emph{Blind Array Access}, which is used to select the next attribute with which to compare 2) \emph{Blind Node Selection}, which selects the next node to traverse to based on the result of the comparison.
The entire protocol is performed using HE, and the comparison is also performed using the functional bootstrapping capability of TFHE. 
Given that only one comparison is performed at each level, this greatly improves performance by not performing unnecessary comparisons. 
The drawback is that since the server does not know which threshold to compare with, it must blindly find the correct threshold with the Blind Array Access subprocedure, which can be computationally expensive. Moreover, the comparison happens between two ciphertexts, as opposed to other approaches which compare encrypted attributes with cleartext thresholds~\cite{tueno, sortinghats}.
For privacy, imbalanced trees are padded with redundant nodes to a full, balanced tree so that all queries are equally expensive. 
The protocol only leaks the depth of the tree and not the number of leaves or any other properties of the tree. 
There is no public implementation of this work available, so we only resort to the theoretical comparison provided in the \Cref{tab:compare-pdte}. 

\paragraph{SortingHats~\cite{sortinghats}}
Cong et al. expanded on the idea of XCMP-style comparisons~\cite{xcmp}.
XCMP focused on comparing two encrypted numbers, but in PDTE, one operand is usually in the clear, making the comparison operation simpler.
Cong et al. proposed a faster comparison operation based on XCMP where only one operand is encrypted and the other is in the clear~\cite{sortinghats}.
We describe this protocol in \Cref{alg:xcmp}.
The operation is done with only two polynomial multiplications, which is much cheaper than comparing two encrypted numbers which requires 16 polynomial multiplications~\cite{xcmp}.
Using their proposed operand, they designed a non-interactive private decision tree evaluation protocol called SortingHats.
The authors used TFHE-based FHE to implement their protocol.
They also proposed using transciphering to reduce communication between the client and server, a method orthogonal to this work to reduce communication costs.
The main limitation of SortingHats is the bit precision, which is currently capped at 11 bits.

\section{Our Comparison Operators}
\label{sec:compare-ops}

In this section, we propose two operators for non-interactive private comparison. First is an extension of the XCMP protocol mentioned in \Cref{sec:background} for larger bit precision.
Second, is a novel protocol based on range covers in combination with constant-weight codes.

\subsection{XXCMP: Extended XCMP Operator}

XCMP only supports the comparison of numbers smaller than $N$. We propose an extension based on XCMP which can support numbers of arbitrary size. We denote this as \emph{XXCMP}.
The idea is to represent large numbers in base $N$. High-order digits are compared first and if they are equal, the next digits are compared. For example, let $a = a_1 N + a_0$ and $b = b_1 N + b_0$ where $a_i,b_i\in[N]$. Then we use the following identity:

\begin{align}
    \mathbb{I}[a>b] = \mathbb{I}[a_1 > b_1] + \mathbb{I}[a_1=b_1] \cdot \mathbb{I}[a_0>b_0] .
\end{align}

For this, we need an equality operator over numbers encoded as XCMP ciphertexts.
Within that algorithm, we use another sub-protocol which checks the equality of two numbers in the XCMP format. This protocol uses the oblivious expansion technique proposed by Angel et al.~\cite{sealpir}. \Cref{alg:extended-xcmp} shows the extension of XCMP for numbers smaller than $N^2$. This algorithm requires one homomorphic multiplication.
This can be extended to numbers with arbitrary length. We have included the algorithm for numbers of arbitrary size in the appendix. In the general case, to compare numbers smaller than $n=N^k$, the multiplicative depth of the circuit is $\ceil{\log_2 k}$ and requires $k(k-1)/2$ homomorphic multiplications.

\begin{algorithm}[ht]
	 \caption{Computing $\mathbb{I}[a > b]$ using Extended XCMP (XXCMP) for $a,b\in[N^2]$ such that $a = a_1 N + a_0$ and $b = b_1 N + b_0$ where $a_i,b_i\in[N]$}
	 \label{alg:extended-xcmp}
	\begin{algorithmic}[1]
        \Procedure{XCMP$_0$}{$\redsymbol{X^a},b$} \Comment{$a,b\in [N]$}
            \State $T \leftarrow -(1 + X + \cdots + X^{N-b-1})$
            \State $R \xleftarrow{\$} R_p$ and $R[0] = 0 \mod p$
            \vspace{1mm}
    	 	\State $\encrypt{C_0} = \redsymbol{X^a} \cdot T + R$
            \vspace{1mm}
 
            \Return $C_0$
        \EndProcedure
        \vspace{3mm}
        \Procedure{XXCMP$_2$}{$\redsymbol{A},b$}
            \Comment{$A\in R_p^2, b\in[N^2]$}
            \State $\redsymbol{X^{a_1}}, \redsymbol{X^{a_0}} \leftarrow \redsymbol{A}$
            \State $\redsymbol{gt_0} \leftarrow$ \textsc{XCMP$_0$}($\redsymbol{X^{a_0}}$, $b_0$)
            \State $\redsymbol{gt_1} \leftarrow$ \textsc{XCMP$_0$}($\redsymbol{X^{a_1}}$, $b_1$)
            \State $\redsymbol{eq_1} \leftarrow$ Oblivious-Expansion($\redsymbol{X^{a_1}}$, $b_1$)
            \State $\redsymbol{C} = \redsymbol{gt_1} + \redsymbol{eq_1} \cdot \redsymbol{gt_0}$
            
            \Return $\redsymbol{C}$
        \EndProcedure
	 \end{algorithmic}
\end{algorithm}

\subsection{Range-Cover Comparison (RCC) Operator}

Assume we have $a,b \in [2^n]$ and we want to compute $\mathbb{I}[a\leq b]$. At a high level, the idea is to use the following statement:
\begin{align}
    a \leq b \iff b \in [a,2^n-1]
\end{align}

which is similar to the RC/PE inclusion problem. However, one end of the interval is always the maximum value. Using this constraint, we define a restricted version of a range cover called the \emph{One-sided Uniform Range Cover (OURC)}. The main difference between a typical range cover and OURC is that an OURC consists of only one prefix node in each level of the prefix tree (including the root level). Hence, the inclusion check requires only $n+1$ equality checks instead of $2n$.
This results in a major performance improvement when running the circuit using HE. \Cref{alg:rcc-compute} shows the procedure for computing the OURC.
The modified OURC/PE inclusion procedure is shown in \Cref{alg:rcc-compute}. We include the procedure for calculating the point encoding as well.

\begin{algorithm}[ht]
	 \caption{\textsc{Calculating OURC and PE}}
	 \label{alg:rcc-compute}
    \begin{algorithmic}[1]
        \Procedure{OURC}{$x,n$}\Comment{$x\in [2^n]$}
            \State $\theta_{OURC} \leftarrow [\texttt{Null}] * (n+1)$
            \If{$x=0$}
                \State $\theta_{OURC}[n] = 0$
                
                \Return $\theta_{OURC}$
            \EndIf
            
            \State $s = 2^n - 1$
            \State $K \leftarrow \{\}$
            \While{$s \geq x$} 
                \State Find $j$ such that $2^j \leq s < 2^{j+1}$
                \State $\theta_{OURC}[j] = 2^{n-j} - 1 - \sum_{k \in K} 2^{k - j}$
                \State $s \leftarrow s - 2^j$
                \State Add $j$ to $K$
            \vspace{2mm}
            \EndWhile
            \Return $\theta_{OURC}$
        \EndProcedure
        \vspace{3mm}
        \Procedure{PE}{$y,n$}\Comment{$y\in [2^n]$}
            \vspace{1mm}
            
            \Return $\left[y, \floor{y/2}, \floor{y/2^2}, \cdots, \floor{y/2^n}\right]$
        \EndProcedure
        \vspace{3mm}
        \Procedure{PE-RC-Inclusion}{$\theta_{OURC}, \theta_{PE}$}
            \State $\theta_{in} = 0$
            \For {$i \in [n+1]$}
                \State $\theta_{in} = \theta_{in} + \mathbb{I}\left[\textsc{OURC}(a,n)[i] == \textsc{PE}(b, n)[i]\right]$
            \vspace{2mm}
            \EndFor
            \Return $\theta_{in}$
        \EndProcedure
    \end{algorithmic}
\end{algorithm}

As shown in Algorithm~\ref{alg:rcc-compute}, $n+1$ equality checks are required for one RCC comparison. We replace the equality checks in each iteration of the for loop with a constant-weight equality operator of Hamming weight $h$. All other operations are additions so the total multiplicative depth of the inequality operator only depends on $h$. Note that the multiplicative depth does not depend on $n$, the bit precision of the numbers.

\begin{figure}[ht]
    \centering
    \includegraphics[width=\columnwidth]{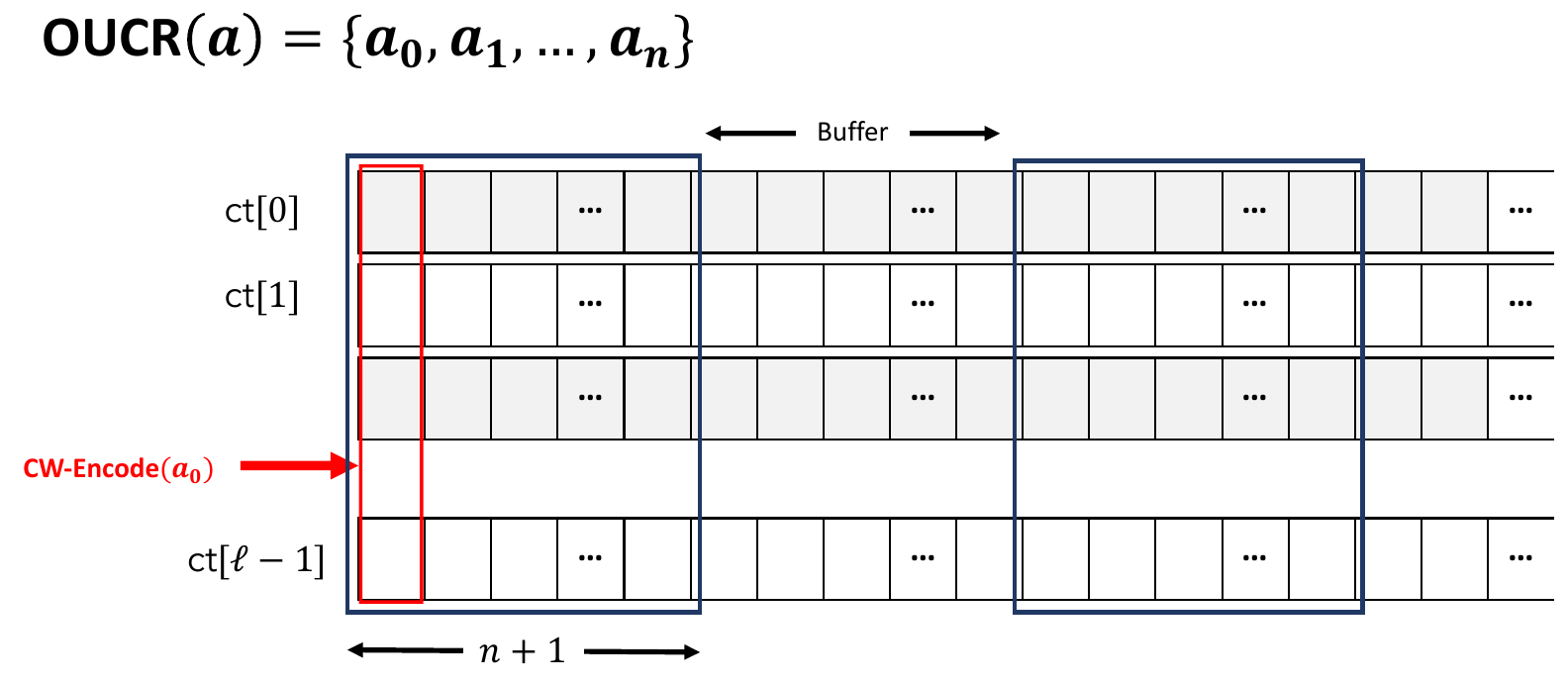}
    \caption{Packing OURC in FV ciphertexts in batched mode.}
    \label{fig:ourc-rep}
\end{figure}

\paragraph{OURC Inclusion using Homomorphic Encryption}
In our private comparison protocol, the client, which holds $a$, sends an encryption of $\textsc{OURC}(a, n)$ to the server. The input is encoded and encrypted such that the comparison is performed efficiently. 
More specifically, if $\textsc{OURC}(a, n) = \{ a_0, a_1, \cdots, a_n \}$, then constant-weight encoding of $a_i$ is spread across $\ell$ ciphertexts. \Cref{fig:ourc-rep} shows a visualization of how the packing happens. Each blue box contains all the bits of information about $\textsc{OURC}(a, n)$.
Using this encoding, each prefix node, $a_i$ occupies exactly $n$ slots in each ciphertext. The number of occupied slots does not depend on the parameters of the constant-weight code, i.e., the Hamming weight.
Multiple pairs of numbers from the client and server can be compared in parallel. 

\Cref{alg:rcc-encode} details the procedure to encode $\textsc{OURC}(a, n)$ across multiple FV ciphertexts in batched mode.
We provide the packing method for the point encoding as well, which is used in the comparison protocol. 
We note that this is not the only method to encode values in FV ciphertexts. We elaborate on other packing methods and why we chose this method in \Cref{sec:discussion}.

\begin{algorithm}[ht]
	 \caption{OURC and PE Encoding}
	 \label{alg:rcc-encode}
	 \begin{algorithmic}[1]
        \Procedure{OURC-Encode}{$a, h, \ell, n$} \Comment{$a\in [2^n]$}
        \State $[a_0, a_1, \cdots, a_n] \leftarrow \textsc{OURC}(a,n)$
        \For{$i\in[n+1]$}
            \State $a'_i = \textsc{CWEncode}(a_i, h, \ell) $ \Comment{$a'_i \in {\mathbb{B}^{\ell}}$}
        \EndFor
        \For{$i \in [\ell]$}
            \State $pt_{OURC}[i] = \left[a'_0[i], a'_1[i], ..., a'_n[i]\right]$
        \vspace{2mm}
        \EndFor
        \Return $pt_{OURC}$
        \EndProcedure
        \vspace{3mm}
        \Procedure{PE-Encode}{$b, h, \ell, n$} \Comment{$b\in [2^n]$}
            \State $[b_0, b_1, \cdots, b_n] \leftarrow \textsc{PE}(b,n)$
            \For{$i\in[n+1]$}
                \State $b'_i = \textsc{CWEncode}(b_i, h, \ell) $ \Comment{$b'_i \in {\mathbb{B}^{\ell}}$}
            \EndFor
            \For{$i \in [\ell]$}
                \State $pt[i] = \left[b'_0[i], b'_1[i], ..., b'_n[i]\right]$
            \vspace{2mm}
            \EndFor
        \Return $pt_{PE}$
        \EndProcedure
	 \end{algorithmic}
\end{algorithm}

Given the encryption of $\textsc{OURC}(a,n)$, we can now outline the procedure for comparison. \Cref{alg:rcc-compare} shows this procedure. The red symbols show the values that are encrypted when the procedure is performed between a client and server.

\begin{algorithm}[ht]
	 \caption{\textsc{RCC Comparison}}
	 \label{alg:rcc-compare}
    \begin{algorithmic}[1]
        \Procedure{RCC-Compare}{$a,b$}
            \State $\redsymbol{a_{ct}} \leftarrow \textsc{OURC-Encode}(a, h, \ell, n)$ \Comment{Done by client}
            \State $b_{pt} \leftarrow \textsc{PE-Encode}(b, h, \ell, n)$
            \State $\redsymbol{\theta} = \textsc{Arith-CW-Eq-Op}(\redsymbol{a_{ct}}, b_{pt})$
            \State $\redsymbol{\theta_{\text{sum}}} \leftarrow \sum_{i=0}^{n} \texttt{Rotate}_i(\redsymbol{\theta}) $
            \State $M\leftarrow 0^N$, $M[0]=1$
            \Comment{Mask}
            \State $\redsymbol{\theta_{\text{cmp}}} \leftarrow \redsymbol{\theta_{\text{sum}}} \otimes M$
            
            \Return $\redsymbol{\theta_{\text{cmp}}}$
        \EndProcedure
    \end{algorithmic}
\end{algorithm}

\section{PDTE using Leveled HE}
\label{sec:levelled-pdte-constructs}
This section describes our two proposed protocols, \xxcmppdte and \rccpdtenospace. We describe the setup, security model, and details about the two protocols.

\subsection{Setup and Security Model}
Our PDTE protocols work in the client/server model.
The server holds a decision tree model, and the client holds a vector of attributes.
The goal is for the client to learn nothing about the server's model other than the inference result. The server should learn nothing from the protocol. Our protocol is non-interactive, i.e., the client uploads its query to the server and waits for a response. The client need not be online while the server processes the query.

Similar to prior work, we work in the semi-honest model, where both the client and server follow the protocol but may try to infer extra information.

Both protocols use homomorphic encryption and assume cryptographic keys such as public keys, evaluation keys, and relinearization keys have been exchanged between the client and server, similar to previous works.

\subsection{XXCMP-PDTE: XXCMP + SumPath}
In \xxcmppdtenospace, we combine the XXCMP comparison with the SumPath algorithm.
This protocol is implemented using FV ciphertexts in polynomial mode.
The protocol has two main parts 1) For each node, compare the correct client attribute to the corresponding node threshold. 2) Run SumPath and get one encrypted result per leaf.
\Cref{alg:xxcmp-pdte} depicts this algorithm. $d.\text{left}$ and $d.\text{right}$ denote the left and right exiting edge from $d$, respectively.
The final result can be compressing the leaves into the same ciphertext. Each leaf can occupy one of the coefficients of the final ciphertext.

Variables that are coloured red are encrypted when running the procedure between a server and client.

\begin{algorithm}[ht]
	 \caption{\xxcmppdte: PDTE using XXCMP}
	 \label{alg:xxcmp-pdte}
	 \begin{algorithmic}[1]
        \Procedure{XXCMP-PDTE}{$\redsymbol{\textbf{x}}, \mathcal{M}$}
            \State $ (\tree, \textbf{a}, \textbf{t}, \textbf{v}) \leftarrow \mathcal{M}$
            \For{$d \in \internal$}
                \State $\redsymbol{c} \leftarrow \textsc{XXCMP}(\redsymbol{\textbf{x}[}\textbf{a}[d]\redsymbol{]}, \textbf{t}[d])$
                \State $d.\redsymbol{\text{left}} \leftarrow \redsymbol{c}$
                \State $d.\redsymbol{\text{right}} \leftarrow 1 - \redsymbol{c}$
            \EndFor
            \For{$\ell \in \leaves$}
                \State $\redsymbol{s(\ell)}= \text{Sum of \redsymbol{edges} from root to } \ell$ 
                \State $r_x,r_y \xleftarrow{\$} \ZZ_p$
                \State $\redsymbol{x(\ell)} \leftarrow r_x \cdot \redsymbol{s(\ell)}$
                \State $\redsymbol{y(\ell)} \leftarrow r_y \cdot \redsymbol{s(\ell)} + v(\ell)$
                \vspace{2mm}
            \EndFor
            \Return $\{(\redsymbol{x(\ell)}, \redsymbol{y(\ell)})\}_{\ell\in \leaves}$
        \EndProcedure
	 \end{algorithmic}
\end{algorithm}

\subsection{RCC-PDTE: RCC + SumPath}
In \rccpdtenospace, comparisons are performed using RCC and the tree is traversed using SumPath. One important difference between this protocol and \xxcmppdte is that all comparisons happen in parallel using the batched mode of FV. The results of the comparisons will occupy different slots of the ciphertext.
\Cref{alg:rcc-pdte} outlines this procedure. In the implementation, we also batched the final result into one ciphertext to reduce communication costs.

\begin{algorithm}[ht]
	 \caption{\rccpdte: PDTE using RCC}
	 \label{alg:rcc-pdte}
	 \begin{algorithmic}[1]
        \Procedure{RCC-PDTE}{$\redsymbol{\textbf{x}}, \mathcal{M}$}
            \State $\tree, \textbf{a}, \textbf{t}, \textbf{v} \leftarrow \mathcal{M}$
            \State $\textbf{t}' \leftarrow $ array of thresholds, aligned with client attributes 
            \State $\redsymbol{c} \leftarrow \textsc{RCC-Compare}(\redsymbol{\textbf{x}}, \textbf{t}')$
            \For{$d \in \internal$}
                \State $\redsymbol{c'}\leftarrow$  Rotate $\redsymbol{c}$ to get $d$ to first slot
                \State $d.\redsymbol{\text{left}} \leftarrow 1 - \redsymbol{c'}$
                \State $d.\redsymbol{\text{right}} \leftarrow \redsymbol{c'}$
            \EndFor
            \For{$\ell \in \leaves$}
                \State $\redsymbol{s(\ell)}= \text{Sum of \redsymbol{\text{edges}} from root to } \ell$ 
                \State $r_x,r_y \xleftarrow{\$} \ZZ_p$
                \State $\redsymbol{x(\ell)} \leftarrow r_x \cdot \redsymbol{s(\ell)}$
                \State $\redsymbol{y(\ell)} \leftarrow r_y \cdot \redsymbol{s(\ell)} + v(\ell)$
                \vspace{3mm}    
            \EndFor    
            \Return $\{(\redsymbol{x(\ell)}, \redsymbol{y(\ell)})\}_{\ell\in \leaves}$
        \EndProcedure
	 \end{algorithmic}
\end{algorithm}

\paragraph{Choosing the Hamming weight}
The Hamming weight used for the constant-weight code directly affects the code length. More specifically, the code length is the smallest $\ell$ such that

\begin{align}
    \binom{\ell}{h} \geq 2^{n}
\end{align}

For $1 \leq h \leq n/2 $, the code length decreases as the Hamming weight increases.
But for $h > n/2$, the code length increases as the Hamming weight increases.
So we do not choose the Hamming weight to be larger than half the bit precision.
Given that analysis, the choice of the Hamming weight requires consideration of the trade-off between runtime and communication costs.
In our evaluation, we plot several Hamming weights to show the effect.

\section{Evaluation}
\label{sec:evaluation}

In this section, we present our evaluation in two parts.
In \Cref{sec:non-interactive-cmp-eval} we benchmark the runtime of our proposed non-interactive private comparison operators in comparison with existing operators.
In \Cref{sec:pdte-eval}, we evaluate PDTE algorithms over decision trees trained over UCI datasets.
We perform ablation studies to measure the performance of different algorithms with respect to precision, the number of client attributes and the size of the decision tree.

\subsection{Benchmarking Private Comparison}
\label{sec:non-interactive-cmp-eval}

\begin{figure}[h!]
    \centering
    \includegraphics[width=\columnwidth]{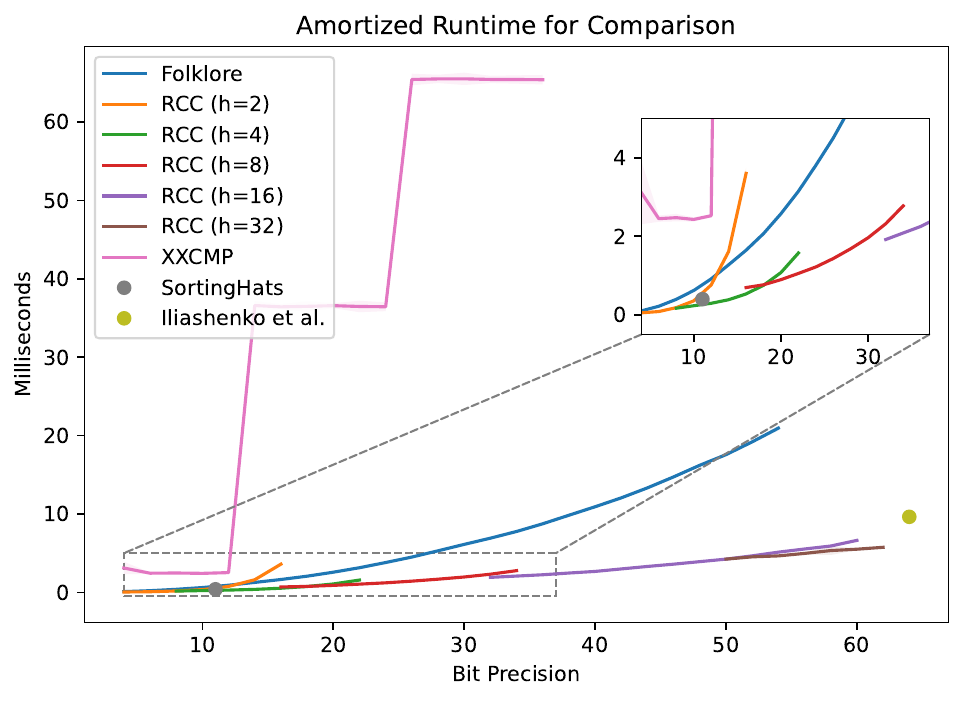}
    \caption{Amortized Time for Comparison. The shaded areas indicate one standard deviation of error. 
    Time for a SortingHats comparison is measured by using the benchmarks in their repository. For RCC, at each point, we use the Hamming weight which has the smallest runtime.
    }
    \label{fig:cmp-benchmark}
\end{figure}

In this experiment scenario, we assume a client wants to compare its input values with that of the server using a private comparison operator.
We measure the computation time for the operators proposed in \Cref{sec:compare-ops}. Specifically, we benchmark 1) RCC 2) Folklore Comparison 3) XXCMP 4) SortingHats Comparison and 5) the work of Iliashenko et al.~\cite{iliashenko2021faster}. The first three are implemented using Microsoft SEAL~\cite{sealcrypto}.
In \Cref{fig:cmp-benchmark} we plot the amortized runtime and communication as a function of the bitlength of the values.
RCC is parameterized by the Hamming weight $h$ which has a significant effect on the performance.
We plot RCC for multiple Hamming weights to show the effect of the parameter.

All experiments are performed 10 times and the average results are reported. The shaded areas show one standard deviation of error.

XCMP and SortingHats perform one comparison at a time, whereas Folklore and RCC are batched and perform the comparison in a SIMD fashion. The input is encoded in the same format as that described in \Cref{fig:ourc-rep}. In this encoding, the number of parallel comparisons is a function of the precision and the parameters of the encryption scheme. \Cref{tab:num-comparisons} shows the number of comparisons for different precisions. The number of parallel comparisons decreases as the precision increases.
Due to our encoding method, the number of parallel comparisons does not depend on the Hamming weight, in the case of RCC.
The work of Iliashenko et al. also performs comparison in a batched manner. We use the univariate variant of their protocol which, per their experiments, produces the best results.

\begin{table}[H]
    \centering
    \begin{tabular}{c|c|c|c|c|c|c|c|c}\toprule
    Bitlength & 8 & 12 & 16 & 20 & 24 & 28 & 32 & 36 \\ \midrule
    \twoline{\# of}{Comps} & 963 & 655 & 496 & 399 & 334 & 287 & 252 & 224 \\
    \bottomrule
    \end{tabular}
    \caption{Number of parallel comparisons in RCC and Folklore comparison as a function of the bit precision.}
    \label{tab:num-comparisons}
\end{table}

\Cref{fig:cmp-benchmark} shows that RCC has the smallest amortized runtime for a private comparison for any bit precision. For a  bit precision of 11, SortingHats and RCC have an approximately similar runtime, but SortingHats can not extend to higher bit precision.

\newcommand{\mysize}[0]{1}
\newcommand{\sizee}[0]{0.24}
\begin{figure*}[ht]
    \centering
    \subfloat[Heart Dataset]{
        \begin{minipage}{\sizee\textwidth}
            \centering
            \includegraphics[width=\mysize\textwidth]{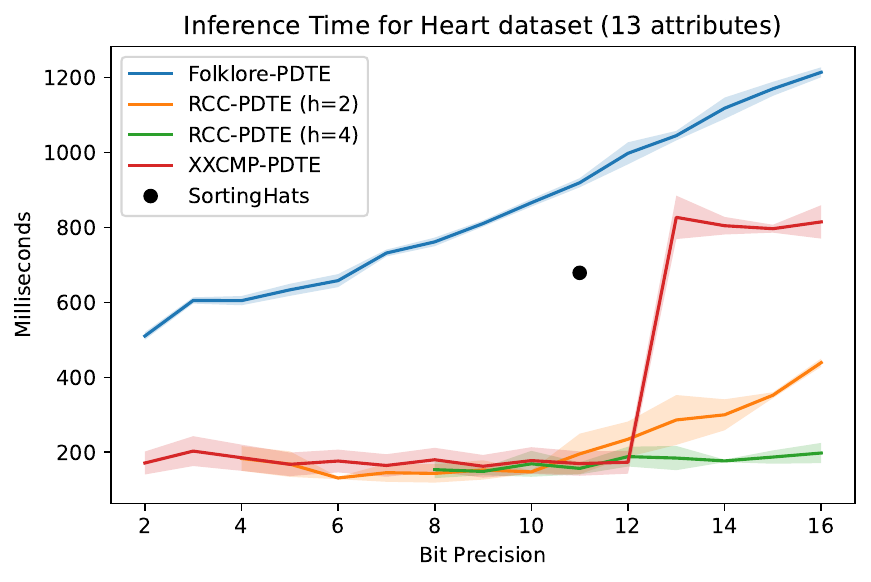}\\
            \includegraphics[width=\mysize\textwidth]{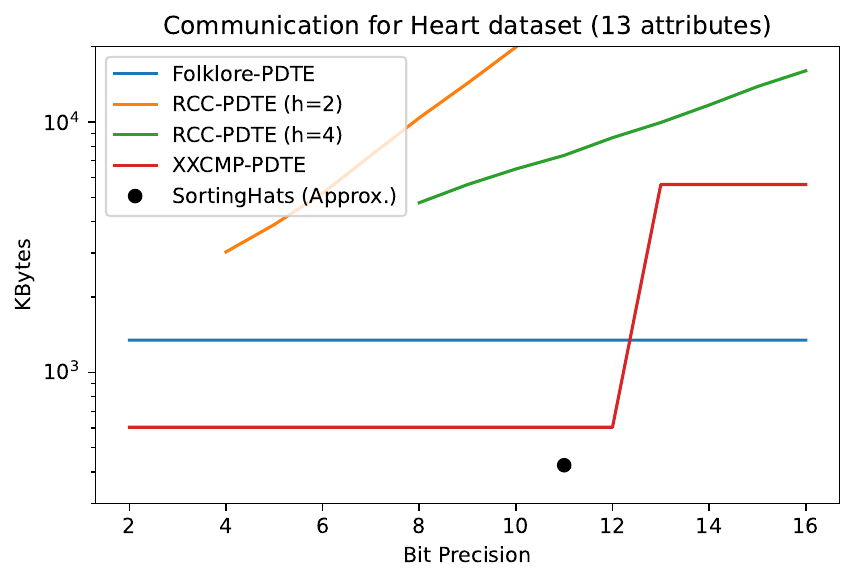}
        \end{minipage}
    }
    \subfloat[Breast Dataset]{
        \begin{minipage}{\sizee\textwidth}
            \centering
            \includegraphics[width=\mysize\textwidth]{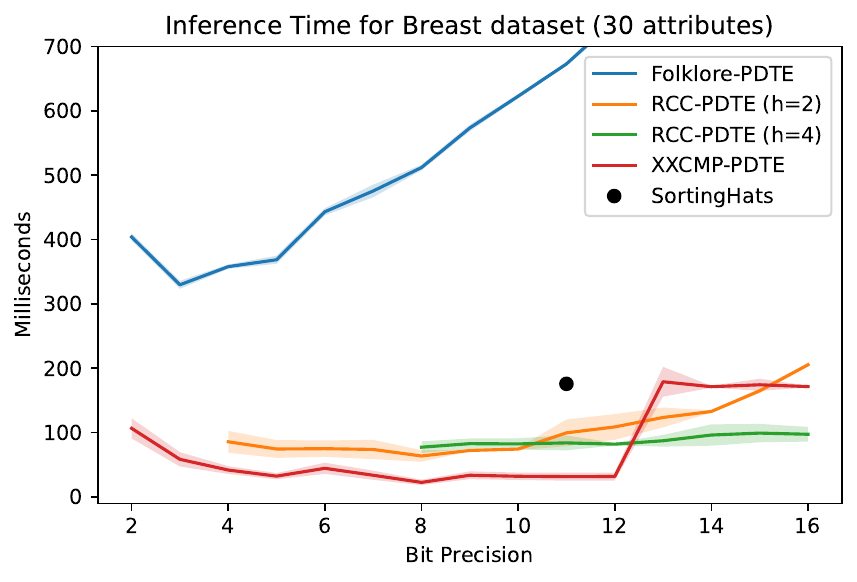}\\
            \includegraphics[width=\mysize\textwidth]{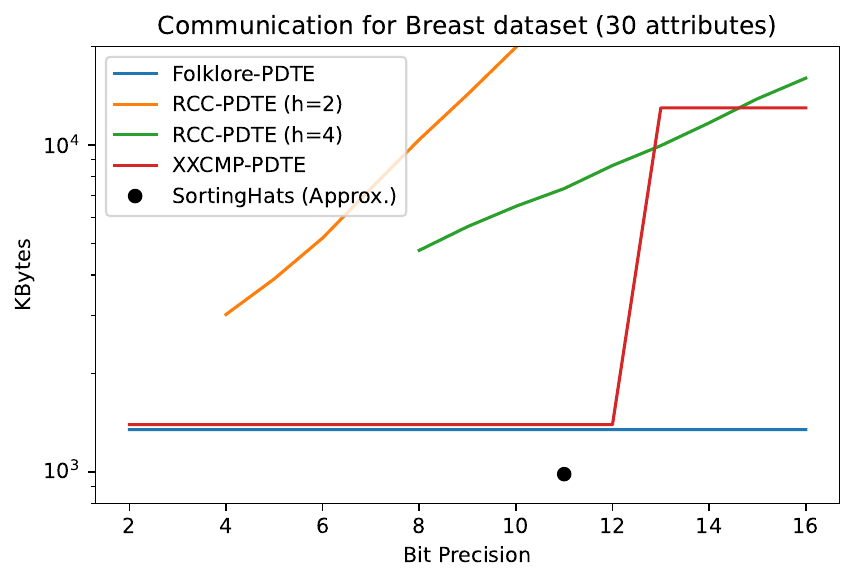}
        \end{minipage}
    }
    \subfloat[Steel Dataset]{
        \begin{minipage}{\sizee\textwidth}
            \centering
            \includegraphics[width=\mysize\textwidth]{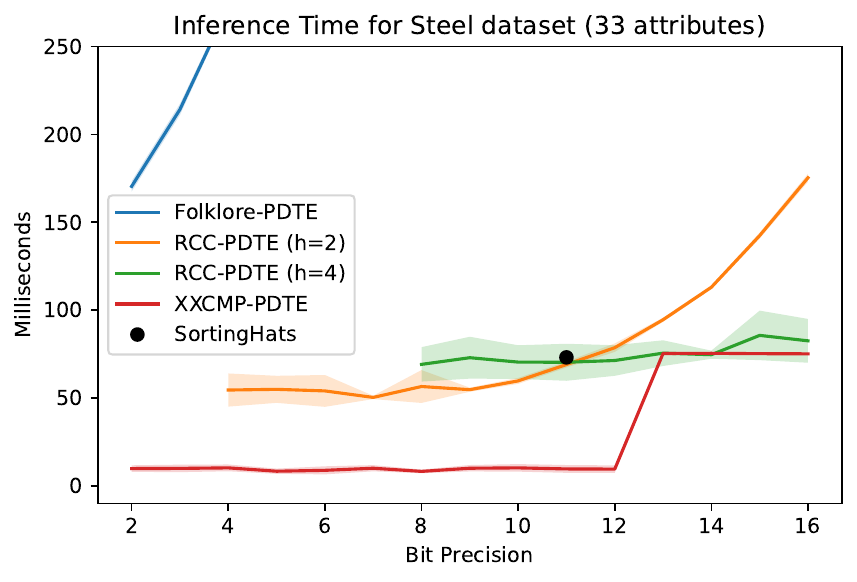}\\
            \includegraphics[width=\mysize\textwidth]{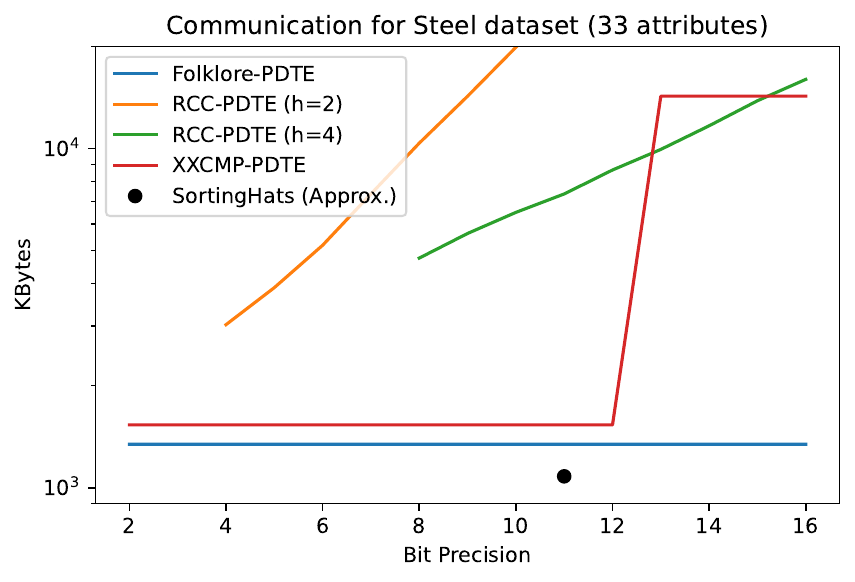}
        \end{minipage}
    }
    \subfloat[Spam Dataset]{
        \begin{minipage}{\sizee\textwidth}
            \centering
            \includegraphics[width=\mysize\textwidth]{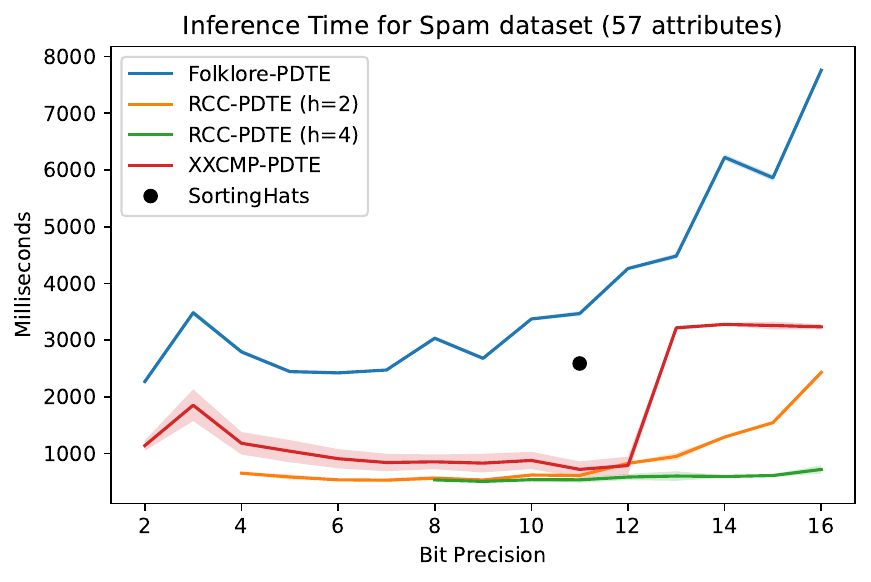}\\
            \includegraphics[width=\mysize\textwidth]{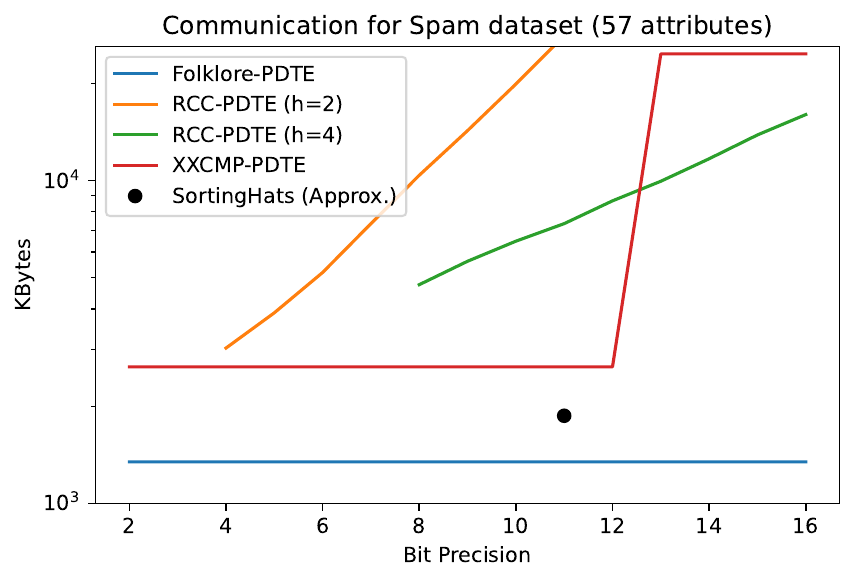}
        \end{minipage}
    }
    \caption{Runtime and Communication for Private Decision Tree Evaluation over four datasets. For each dataset, the left graph plots the runtime, and the right graph plots the communication. The shaded area shows one standard deviation of error.}
    \label{fig:datasets-eval}
\end{figure*}

\subsection{Benchmarking PDTE}
\label{sec:pdte-eval}

In this subsection, we compare the proposed private decision tree evaluation protocols in terms of server computation time and communication between the client and server.
We first benchmark the performance whilst inferring decision trees trained on common UCI datasets and measure the communication and computation overhead.
Our experiments over datasets show that the performance of these algorithms is tied to the number of client attributes and the size of the decision tree. Hence, we ablate with respect to these two parameters to understand the effect.

\subsubsection{Implementation \& Experimental Details}
\xxcmppdte and \rccpdte are implemented as described in \Cref{sec:levelled-pdte-constructs}. We parallelized some steps in both protocols to enhance performance. Particularly, in \rccpdtenospace, the constant-weight equality operator was shown to be highly parallelizable by Mahdavi and Kerschbaum~\cite{cpir}.
We also implement a baseline solution which we denote as \emph{Folklore-PDTE}. 
Folklore-PDTE is implemented similarly to \rccpdtenospace, with only the comparison replaced with the Folklore comparison from \Cref{sec:background-non-interactive-cmp}.
All three algorithms are implemented using Microsoft SEAL version 4.0\footnote{\url{https://github.com/microsoft/SEAL}}, which implements the FV cryptosystem in polynomial and batching mode~\cite{sealcrypto}.
Our implementation is publicly available on Github.\footnote{\url{https://github.com/RasoulAM/private-decision-tree-evaluation}}

For SortingHats, we use the implementation provided by the authors\footnote{\url{https://github.com/KULeuven-COSIC/SortingHat}}.
We activate the parallelization flag and use artificial input.
Experiments are conducted on an Intel(R) Xeon(R) Platinum 8368 CPU @ 2.40GHz server running Ubuntu 22.02 with 32 cores.
All experiments are repeated 10 times and the average is reported. The shaded areas indicate the standard deviation of the measurements.

\subsubsection{Evaluation over Datasets.}
We train decision trees over four datasets from the UCI repository~\cite{Dua:2019}, Heart, Breast, Spam, and Steel, which are also used in related work~\cite{tueno, sortinghats}. We train decision trees with the desired precision using the Concrete-ML framework~\cite{ConcreteML}. \Cref{tab:datasets} shows the properties of the datasets.
The structure of the decision tree changes as the precision increases.
In general, a decision tree with higher precision has fewer nodes.

\begin{table}[H]
    \centering
    \begin{tabular}{c|c|c|c}
    \toprule
        Name & ID & \twoline{\# of}{Classes} & \twoline{\# of}{Attributes}\\
        \midrule
        Breast & 1510 & 2 & 30 \\
        Steel  & 1504 & 2 & 33 \\
        Heart  & 1565 & 5 & 13 \\
        Spam   &   44 & 2 & 57 \\
        \bottomrule
    \end{tabular}
    \caption{Characteristics of UCI datasets used in our evaluation}
    \label{tab:datasets}
\end{table}

In \Cref{fig:datasets-eval}, communication and computation are plotted as a function of the precision for Folklore-PDTE, \xxcmppdtenospace, and \rccpdtenospace. SortingHats does not permit arbitrary precision, so we plot it as one point in the graphs.

\Cref{fig:datasets-eval} shows the results for four datasets. Folklore-PDTE is consistently the slowest of all solutions. 
None of the benchmarked approaches is consistently better, but in all cases, it is either \xxcmppdte or \rccpdtenospace.
In communication, Folklore-PDTE is dominant given that it has the most compact representation for a number, but given its impractical runtime, we can dismiss that. Hence, if we disregard that, SortingHats has the least communication overhead compared to \xxcmppdte and \rccpdtenospace.

These experiments show that there is not a dominant solution that wins in all cases for all metrics. Communication and computation are a function of many factors, such as the number of client attributes and the number of decision nodes. We perform ablations to better understand the effect of each of these parameters.

\subsubsection{Ablation over Number of Attributes.}
In this experiment, we benchmark PDTE over a synthetic decision tree with a varying number of client attributes. Specifically, we generate a synthetic balanced tree of depth 6 (with 31 decision nodes) with three different bit precision, $n=8,16,26$. Note that the same results hold for trees of other sizes and shapes (balanced or unbalanced). 
We plot the communication complexity of \rccpdte and \xxcmppdte as we vary the number of client attributes from 5 to 100.

\begin{figure}[H]
    \centering
    \includegraphics[width=0.4\textwidth]{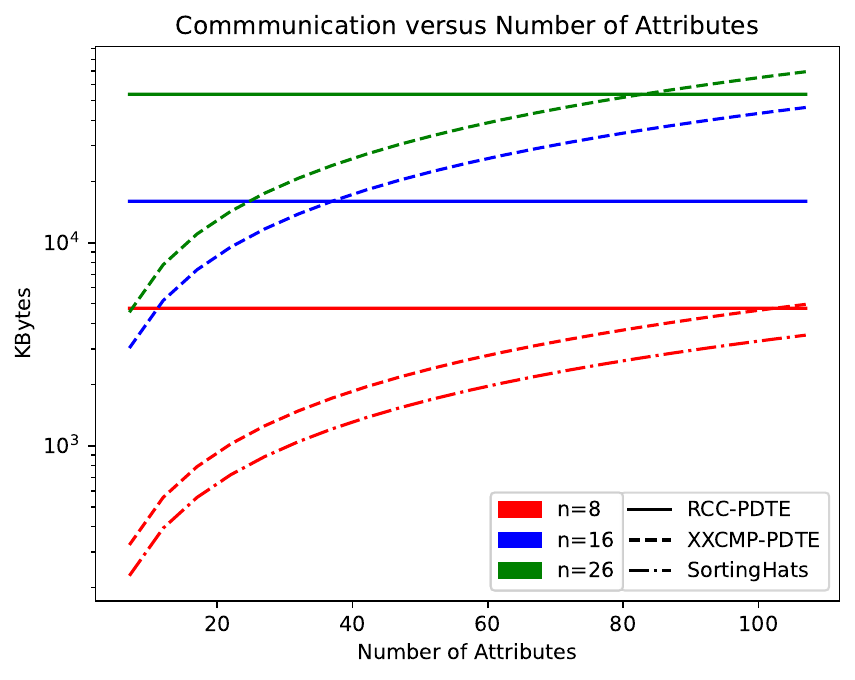}
    \caption{Communication cost of PDTE as a function of the number of attributes.}
    \label{fig:comm-num-attr}
\end{figure}

SortingHats and \xxcmppdte encrypt each client attribute in a separate RLWE ciphertext.
Hence, the communication complexity scales linearly with the number of client attributes (notice the logarithmic vertical axis). 
For a given precision, \rccpdte has constant communication cost due to the use of batch encoding.
When the number of client attributes is small, \rccpdte has a higher communication cost compared to other solutions.
SortingHats has the smallest communication overhead when precision is less than 11 bits. For precision higher than 11 bits, \xxcmppdte is the best solution in terms of communication.
While \rccpdte has the highest communication cost when the number of attributes is small, this high overhead shrinks as the number of attributes grows and in some cases, it becomes the dominant solution.
For example, at 16-bit precision, \rccpdte has the least communication overhead once the number of client attributes exceeds 40.

Note that the runtime of both algorithms does not have a noticeable change as the number of attributes changes. Hence, we only report the approximate runtimes of each approach in \Cref{tab:time-num-attr}. Similar to the results shown in \Cref{fig:datasets-eval}, \xxcmppdte is the fastest for a low bit precision, but \rccpdte overtakes it for higher bit precision. SortingHats is only applicable for low precision and is slower in that case.

\begin{table}[H]
    \centering
    \begin{tabular}{c|c|c|c}
        \toprule
        \multirow{2}{*}{Precision (bits)} & \multicolumn{3}{c}{Runtime (ms)} \\
        & SortingHats & \xxcmppdte & \rccpdte \\
        \midrule
         8  & 648 - 662 & 133 - 155   & 149 - 170  \\
        16  &  -      & 673 - 753 & 187 - 234  \\
        26  &  -      & 752 - 845 & 747 - 966  \\
        \bottomrule
    \end{tabular}
    \caption{Approximate runtime for private evaluation of balanced decision tree of depth 6 (with 31 decision nodes). The number of attributes varies from 7-100}
    \label{tab:time-num-attr}
\end{table}

\subsubsection{Ablation over Number of Nodes.}
\begin{figure}[H]
    \centering
    \includegraphics[width=0.4\textwidth]{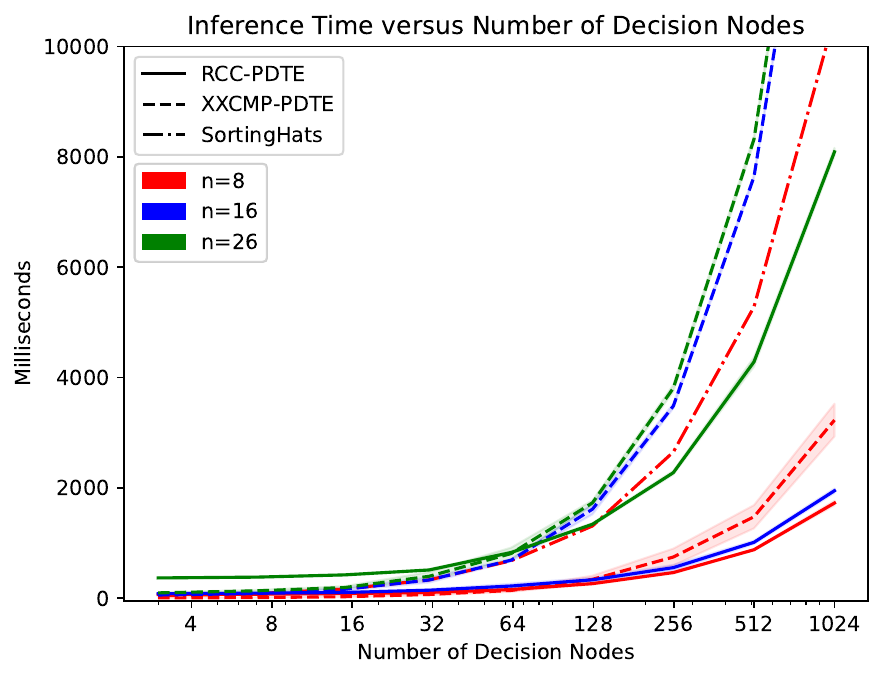}
    \caption{Runtime as a function of the number of nodes.}
    \label{fig:time-num-nodes}
\end{figure}

In this experiment, we benchmark PDTE over synthetic trees as we vary the number of decision nodes. We fix the number of attributes to 32, but the same results hold for a different number of attributes. We generate balanced trees with depths up to 10, but the results do not depend on the shape of the tree, and the same results hold if the tree is unbalanced.
This is because the tree traversal algorithm, SumPath, only takes up at most 10\% of the total runtime. Hence, a change in the shape of the tree does not significantly impact the runtime.
\Cref{fig:time-num-nodes} shows the runtime as a function of the number of nodes for $n=8,16,26$. Note that the horizontal axis is logarithmic.

As expected, the runtime of \xxcmppdte increases linearly with the number of decision nodes, given that one comparison is performed for each decision node in the tree.
Due to batched computations, the runtime of \rccpdte only increases if more ciphertexts are required for the comparisons.
For low precision, \xxcmppdte has better runtime compared to \rccpdtenospace. However, for larger decision trees, \rccpdte is faster.

In all the reported protocols, the communication cost is a function of the precision and number of client attributes (and the Hamming weight, in the case of \rccpdtenospace).
Communication cost does not depend on the size of the decision tree.
Hence, we only report the communication cost of each protocol in \Cref{tab:comm-num-nodes}.

\begin{table}[H]
    \centering
    \begin{tabular}{c|c|c|c}
    \toprule
        \multirow{2}{*}{\twoline{Precision}{(bits)}} & \multicolumn{3}{c}{Communication (KB)} \\
        & SortingHats & \xxcmppdte & \rccpdte  \\
        \midrule
         8 & 2342 & 1486 & 4757     \\
        16 & -   & 13847 & 16001  \\
        26 & -   & 20770 & 53795  \\
    \bottomrule
    \end{tabular}
    \caption{Communication cost for private evaluation of balanced tree over an input with 32 attributes. The depth of the tree to evaluate is up to 7.}
    \label{tab:comm-num-nodes}
\end{table}

\subsection{Summary of Results}

For non-interactive private comparison, XXCMP and RCC can be utilized for arbitrary precision, while SortingHats has limited precision. If a single comparison with arbitrary precision is required, XXCMP is a good option; however, RCC offers a significantly better amortized time for comparing numerous pairs.

In the context of PDTE, for low precision, either SortingHats, \xxcmppdtenospace, or \rccpdte may be faster. The optimal solution is contingent on a combination of factors, including bit precision, the number of decision nodes, and the number of client attributes. For low precision, SortingHats is superior in terms of communication, but \rccpdte and \xxcmppdte are generally faster.

When dealing with bit precision higher than 11, \xxcmppdte and \rccpdte are the only practical available options, with \rccpdte proving to be faster, particularly as the number of decision nodes increases.

\section{Discussion and Future Work}
\label{sec:discussion}

\paragraph{High-precision Applications}
While decision trees may be achievable with smaller precision and quantization-aware training, there are other cases where the precision can not be sacrificed.
For example, in the case of intrusion detection, servers may want to check to see if an IP, which is a 32-bit number, is in a specific range or not.

\paragraph{Extension of SortingHats}
Other protocols such as SortingHats can also be modified to accommodate larger precision. This can be done using a similar algorithm to XXCMP but with a fully homomorphic scheme such as TFHE. Specifically, the equality check that is performed in line 9 of \Cref{alg:extended-xcmp} can be performed with functional bootstrapping instead.

\paragraph{Other Packing Methods}
Homomorphic encryption in the batched setting offers a lot of flexibility to choose the packing method.
Other packing methods can improve communication, particularly when trying to reduce the time for one inference. Tools such as HELayers~\cite{helayers} can help find the other packing strategies.

\paragraph{Comparison with PROBONITE}
There is currently no public implementation for PROBONITE available. We conducted a theoretical comparison of PROBONOTE with other PDTE protocols in \Cref{tab:compare-pdte}, but a practical comparison would also be interesting as part of future work.

\section{Conclusion}

In this work, we propose two protocols for non-interactive private decision tree evaluation leveraging levelled homomorphic encryption, \xxcmppdte and \rccpdtenospace. These protocols are based on XXCMP and RCC, two non-interactive comparison protocols which can efficiently compare numbers of arbitrary precision with a constant multiplicative depth.

Our experimental analysis demonstrates that several protocols can be used when the client's input features a small number of attributes, the decision tree remains small, and lower precision is acceptable.
However, when faced with many client attributes, large decision trees or the necessity for high precision, \xxcmppdte and \rccpdte emerge as better options compared to SortingHats.
In some cases, these two protocols are up to 5 times faster than SortingHats.
In very large decision trees, \rccpdte is the best solution and can infer a decision tree with over 1000 nodes and 16 bits of precision in under 2 seconds.

\bibliographystyle{ACM-Reference-Format}
\bibliography{references}

\appendix

\section{Arithmetic Constant-weight Equality}
\label{sec:cw-stuff}

Mahdavi and Kerschbaum proposed constant-weight equality operators, which were equality operators used to compare constant-weight codes with a constant multiplicative depth. We use these operators as a building block in our work to achieve a PDTE protocol that has a multiplicative depth independent of precision and the depth of the tree.

Mahdavi and Kerschbaum offer a function for encoding numbers as constant-weight codes. We use this function with the added option of encoding a null value. This is useful in our case since we may need to encode null elements as well. Null elements are encoded as the all-zero string. 

\Cref{alg:cw-encode} shows this algorithm.
$CW(\ell,h)$ denotes the set of constant-weight codes with length $\ell$ and Hamming weight $h$.

\begin{algorithm}[ht]
	 \caption[]{\texttt{CW-Encode}~\cite{cpir}}
	 \label{alg:cw-encode}
	 \begin{flushleft}
		 \textbf{Input:} $x \in [2^n]\cup\{\texttt{Null}\}$, $\ell,h \in \NN$ such that $\binom{\ell}{h} \geq 2^n$
	 \end{flushleft}
	 \begin{algorithmic}[1]
        \If {x==\texttt{Null}}
            \State \textbf{return} $0^\ell$
        \EndIf
	 	\State $r \leftarrow x$, $h' \leftarrow h$, $y \leftarrow 0^\ell$
	 	\For {$\ell'  = \ell-1, ..., 1, 0 $}
		 	\If {$r \geq \binom{\ell'}{h'}$} \\
				\hspace{10mm}$y[\ell']=1$ \\
				\hspace{10mm}$r = r - \binom{\ell'}{h'}$ \\
				\hspace{10mm}$h' = h'-1$
			\EndIf
			\If {$h=0$}
			    break
			\EndIf
	 	\EndFor
        \State \textbf{return} y
	 \end{algorithmic}
	 \begin{flushleft}
    	 \textbf{Output:}  $y \in CW(\ell,h) \cup \{0^\ell\}$\\
	 \end{flushleft}
\end{algorithm}

The arithmetic constant-weight equality operator is the same as proposed by Mahdavi and Kerschbaum. The input can now be the all-zero string as well. Comparing anything with the null string will yield a non-match.


\section{XXCMP for Arbitrary-length Numbers}

XXCMP can compare numbers of arbitrary size. \Cref{alg:extended-xcmp-general} shows the general XXCMP algorithm which compares two number $a,b\in[N^k]$, for some known parameter $k$. The output of the comparison is in the constant

\begin{algorithm}[H]
	 \caption{Computing $\mathbb{I}[a>b]$ using Extended XCMP (XXCMP) for $a,b\in[N^k]$ such that $a = (a_{k-1},\cdots,a_1,a_0)_N$ and $b=(b_{k-1},\cdots,b_1,b_0)_N$ where $a_i,b_i\in[N]$}
	 \label{alg:extended-xcmp-general}
	\begin{algorithmic}[1]
        \Procedure{XXCMP}{$\redsymbol{A},b$}
            \Comment{$A\in R_p^2, b\in[N^2]$}
            
            \State $(\redsymbol{X^{a_{k-1}}}, \cdots ,\redsymbol{X^{a_1}}, \redsymbol{X^{a_0}}) \leftarrow \redsymbol{A}$
            \For{$i \in [k]$}
                \State $\redsymbol{gt_i} \leftarrow$ \textsc{XCMP$_0$}($\redsymbol{X^{a_i}}$, $b_i$)
            \EndFor
            \For{$i \in [k]$}
                \State $\redsymbol{eq_i} \leftarrow$ Oblivious-Expansion($\redsymbol{X^{a_i}}$, $b_i$)
            \EndFor
            
            \State $\redsymbol{C} = \sum_{i=0}^{k-1} \redsymbol{gt_{i}} \cdot \prod_{j=i+1}^{k-1} \redsymbol{eq_{j}}$
            
            \Return $\redsymbol{C}$
        \EndProcedure
	 \end{algorithmic}
\end{algorithm}

\section{RC/PE Inclusion}
\label{sec:rc-pe-compare}

We can test the relationship between a range cover and point encoding using \Cref{alg:rc-compare}.
Assume we want to check $c\in [a,b]$ using $RC_n(a,b)$ and $PE_n(c)$.

In this notation, $RC_n(a,b)$ contains $2n$ prefix nodes and we assume the prefix nodes from level $i$ of the prefix tree are in $RC_n(a,b)[2i]$ and $RC_n(a,b)[2i+1]$ (and some levels may be empty).
$PE_n(c)$ contains $n$ prefix nodes and the prefix node from level $i$ of the prefix tree is in $PE_n(c)[i]$.
The inclusion algorithm requires $2n$ equality check at most.

\begin{algorithm}[ht]
	 \caption{\textsc{RC/PE Inclusion}}
	 \label{alg:rc-compare}
	 \begin{flushleft}
		 \textbf{Input:} $RC_n(a,b), PE(c,n)$ \\
	 \end{flushleft}
	 \begin{algorithmic}[1]
        \State $\theta_{included} = 0$
        \For {$i \in [n]$}
            \State $\theta_{included} = \theta_{included} + \mathbb{I}\left[RC_n(a,b)[2i] == PE_n(c)[i]\right]$
            \State $\theta_{included} = \theta_{included} + \mathbb{I}\left[RC_n(a,b)[2i+1] == PE_n(c)[i]\right]$
        \EndFor
	 \end{algorithmic}
	 \begin{flushleft}
		 \textbf{Output:} $\mathbb{I}\left[c \in [a,b]\right]$
	 \end{flushleft}
\end{algorithm}

\end{document}